%% file: main.tex
\documentclass[11pt]{article}

\usepackage{enumitem}
\usepackage{booktabs}
\usepackage{multirow}
\usepackage{amsmath}  
\usepackage{graphicx}
\usepackage{amssymb}  
\usepackage{siunitx} 
\usepackage[table]{xcolor} 
\usepackage{float}
\usepackage{tabularx}   
\usepackage{caption}
\usepackage{tcolorbox}
\usepackage{afterpage} 
\usepackage{CJKutf8}
\usepackage{array}
\usepackage{arydshln}

\usepackage[final]{acl}

\usepackage{times}
\usepackage{latexsym}

\usepackage[T1]{fontenc}

\usepackage[utf8]{inputenc}

\usepackage{microtype}
\usepackage{inconsolata}

\title{MCSC-Bench: Multimodal Context-to-Script Creation \\for Realistic Video Production}
  
\author{
    Huanran Hu$^{1}$ \thanks{Equal contribution.}, Zihui Ren$^{2}$ \footnotemark[1], Dingyi Yang$^{3}$ \footnotemark[1], Liangyu Chen$^1$,
    Qixiang Gao$^2$, Tiezheng Ge$^2$, Qin Jin$^{1}$ \thanks{Corresponding author.}
    \vspace{2mm} \\
    $^1$Renmin University of China \quad $^2$Alibaba Group \quad $^3$Nanyang Technological University \\
    \vspace{1mm} \\
    {\tt\small \{huanranhu, liangyuchen, qjin\}@ruc.edu.cn} \\
    {\tt\small \{renzihui.rzh, gaoqixiang.gqx, tiezheng.gtz\}@taobao.com} \\
    {\tt\small dingyi.yang@ntu.edu.sg}
}

\begin{document}

\maketitle

\begin{abstract}

Real-world video creation often involves a complex reasoning workflow of selecting relevant shots from noisy materials, planning missing shots for narrative completeness, and organizing them into coherent storylines. However, existing benchmarks focus on isolated sub-tasks and lack support for evaluating this full process. To address this gap, we propose \emph{Multimodal Context-to-Script Creation (MCSC)}, a new task that transforms noisy multimodal inputs and user instructions into structured, executable video scripts. We further introduce \textbf{MCSC-Bench}, the first large-scale MCSC dataset, comprising 11K+ well-annotated videos. Each sample includes: (1) redundant multimodal materials and user instructions; (2) a coherent, production-ready script containing material-based shots, newly planned shots (with shooting instructions), and shot-aligned voiceovers. MCSC-Bench supports comprehensive evaluation across material selection, narrative planning, and conditioned script generation, and includes both in-domain and out-of-domain test sets. Experiments show that current multimodal LLMs struggle with structure-aware reasoning under long contexts, highlighting the challenges posed by our benchmark. Models trained on MCSC-Bench achieve SOTA performance, with an 8B model surpassing Gemini-2.5-Pro, and generalize to out-of-domain scenarios. Downstream video generation guided by the generated scripts further validates the practical value of MCSC. Datasets will be public soon.

\end{abstract}

\input{section/1_introduction}

\input{section/3_task}

\input{section/4_dataset}

\input{section/5_1_experiment}

\input{section/5_2_capacity}

\input{section/5_3_multi_agent}
\input{section/5_4_ablation}
\input{section/5_5_generation}

\input{section/2_related_work}

\input{section/6_conclusion}

\bibliography{refer}

\appendix

\section{Appendix A: Implementation and Dataset Details}
\label{sec:appendix}
This section provides comprehensive details regarding the dataset construction, model prompting strategies, and human evaluation protocols to ensure the reproducibility of our work.
\input{section/Appendix/anno}
\input{section/Appendix/prompt}
\input{section/Appendix/train}

\section{Appendix B: Extended Evaluation Protocols}
To provide a multifaceted assessment of model performance, we detail the traditional metrics and the setup for human evaluation.

\input{section/Appendix/bleu}
\input{section/Appendix/eval}

\section{Appendix C: Supplementary Experiments and Analysis}
This section presents extended experimental results.

\input{section/Appendix/multi_agent}
\input{section/Appendix/case_study}
\input{section/Appendix/video_generation}

\input{section/Appendix/image}

\end{document}

%% file: section/1_introduction.tex
\section{Introduction} 

\begin{figure*}[t]
    \centering
    \includegraphics[width=\linewidth]{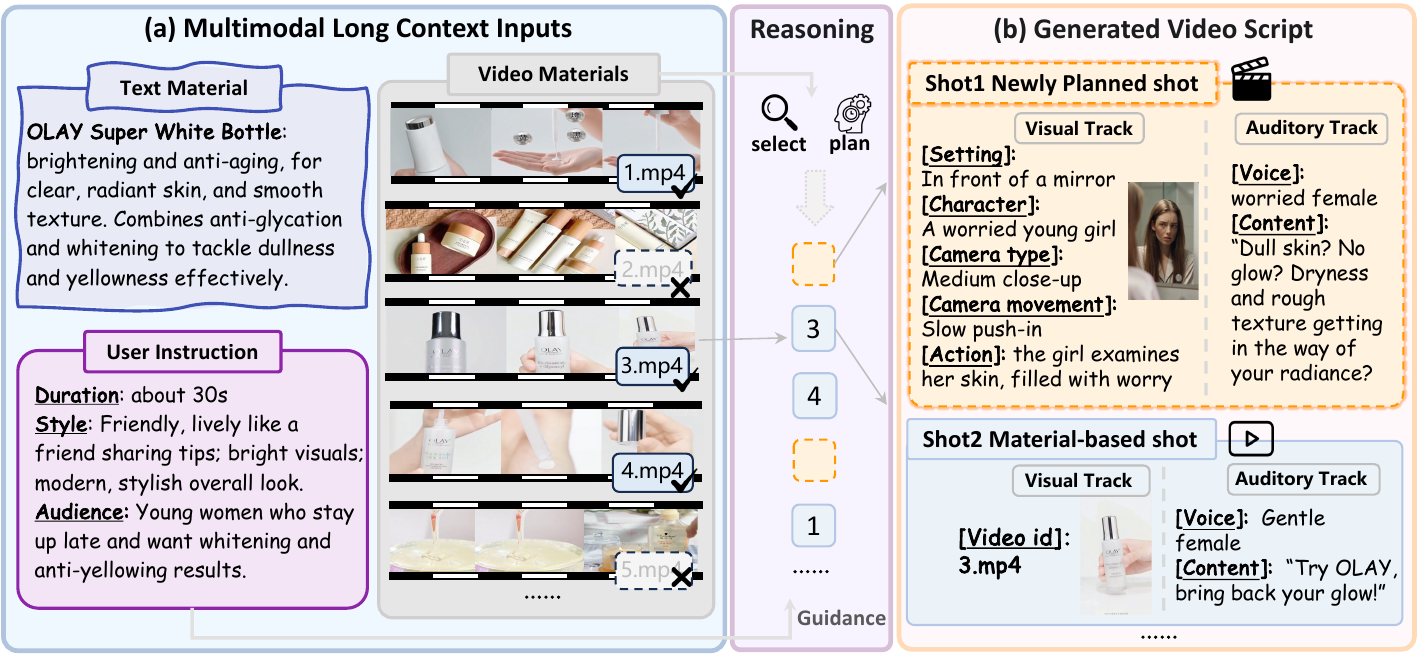}
    \vspace{-8pt}
    \caption{An overview of our \textbf{Multimodal Context-to-Script Creation (MCSC) } task. Models should comprehend the multimodal long contexts, create the plot, and output the structured script, which includes material-based shots and newly planned shots.}
    \label{fig:task}
\end{figure*}

Automated video creation has become an important tool for modern content production across advertising, entertainment, and social media~\citep{chi2020automatic, zhu2023moviefactory, zheng2024opensora, henschel2025t2v, mu2026script}. While recent advances in video creation have shown promising results, most existing systems assume simplified settings, such as relying on short text prompts or pre-curated multimodal inputs ~\citep{hong2022cogvideo, kondratyuk2023videopoet, he2023animate, chen2024hollmwood,qian2025vc-llm}, which do not reflect real-world creative workflows. 

As illustrated in Figure~\ref{fig:task}, real-world video creation typically begins with pre-existing textual materials and redundant video materials \cite{dancyger2018technique, pearlman2018documentary, wang2019write}. Guided by specific user instructions, creators must identify useful video shots, discard irrelevant content, and plan additional shots to bridge narrative gaps~\cite{bourgeois2014creativity,zhang2022ai,li2025fromshots}. This process makes it a long-context multimodal reasoning and structured planning problem, rather than a pure generation problem \cite{wang2024lave,ji2022vscript, han2023shot2story20k, chen2024hollmwood,dong2026vista}. We argue that an \textit{executable video script} serves as a critical intermediate representation between noisy multimodal contexts and the final video. As shown in Figure \ref{fig:task} (b), such scripts weave existing shots and newly planned shots into a coherent sequence, provide shooting instructions for new shots, and include aligned voiceovers for each shot.

Driven by this insight, we introduce \textbf{Multimodal Context-to-Script Creation (MCSC)} (Figure~\ref{fig:task}), a new task that formalizes structured video script production. 
Given multimodal materials and user instructions, models are required to: (i) select relevant content from redundant inputs; (ii) plan additional shots for narrative completeness; and (iii) generate a structured script.
The production-ready script enables controllable planning and supports downstream video generation.

{Building on this formulation, we present \textbf{MCSC-Bench}, the first large-scale dataset for MCSC. It comprises over 11K professionally annotated samples. Each sample includes: (i)~a set of input video shots including relevant and distractor materials; (ii)~text materials; (iii)~user instructions; and (iv)~structured output scripts containing shooting instructions for newly planned shots, material labels for selected shots, and aligned voiceovers. 
To enable systematic evaluation, we design both automatic rule-based metrics and multi-dimensional metrics (detailed in Section~\ref{sec:multi_aspect}) to assess instruction following, material utilization, and script quality. Compared with existing benchmarks, MCSC-Bench jointly evaluates \textit{multimodal comprehension}, \textit{structure-aware planning}, and \textit{conditioned narrative generation}, making it a challenging and practical benchmark for long-context multimodal reasoning and controllable content creation.

Extensive experiments on MCSC-Bench reveal that existing multimodal LLMs, especially open-source models, still struggle with multimodal long-context reasoning and subsequent script generation, highlighting the benchmark’s discriminative power and challenge. We find that explicit task decomposition helps reduce the complexity of MCSC. Models trained on MCSC-Bench via SFT and RL achieve strong performance, with an 8B model surpassing Gemini-2.5-Pro. Finally, downstream video generation demonstrates that the produced scripts effectively support coherent and high-quality video creation, highlighting the practical value of MCSC-Bench for automated video production.}

The main contributions of this work are threefold: 
(1) { \textbf{MCSC}, a novel and practical task to advance automatic video creation, which requires models to select relevant shots from redundant multimodal materials, plan necessary new shots, and organize them into a coherent, structured script.}
(2) \textbf{MCSC-Bench}, { the first large-scale dataset for MCSC, comprising 11K+ well-annotated samples. We design rule-based and multi-dimensional metrics to enable systematic evaluation of current MLLMs.}
(3) Extensive experiments demonstrate that MCSC-Bench effectively benchmarks current MLLMs and reveals their cognitive bottlenecks in MCSC. Models fine-tuned on our dataset achieve SOTA performance, and downstream video generation experiments validate its practical value for real-world video production.

%% file: section/3_task.tex
\section{MCSC: Task Design and Evaluation}
\label{sec:task_design}

\input{tables/tab1_compare}

To better reflect real-world video script production processes, we propose a novel task: \textbf{Multimodal Context-to-Script Creation (MCSC)}. This section introduces the task formulation and the corresponding evaluation metrics.

\subsection{Task Definition}
The input to the MCSC task is a heterogeneous multimodal context $\mathcal{C} = (\mathcal{V}, \mathcal{T}, \mathcal{I})$, where each component provides complementary information for script construction (Figure~\ref{fig:task}).
$\mathcal{V} = \{v_1, v_2, \dots, v_M\}$ denotes a set of $M$ raw video shots, comprising \textbf{Relevant Materials} ($\mathcal{V}_{a}$) sourced from related videos, and \textbf{Distractor Materials} ($\mathcal{V}_{d}$) from irrelevant videos.
$\mathcal{T}$ is a collection of \textbf{Text Materials}, such as core theme, product descriptions, and brand information.
$\mathcal{I}$ is the \textbf{User Instruction}, specifying constraints such as target duration, stylistic preferences, and intended audience.

Given the multimodal context $\mathcal{C}$, the core objective of MCSC is to generate a structured, coherent, and engaging script $\mathcal{S}$, 
represented as
an ordered sequence of $N$ shots: $\mathcal{S} = (s_1, \dots, s_N)$. Each shot $s_i$ contains both a visual and an auditory track:
\begin{itemize}[leftmargin=*, itemsep=0pt, topsep=0pt]
\item \textbf{Visual Track} specifies the shot's visual content in two forms: (i) Material-based Shot ($S_{\text{mat}}$), a shot directly selected from the input material set $\mathcal{V}$; and (ii) Newly Planned Shot ($S_{\text{new}}$), a newly created scene described textually in terms of setting, characters, camera actions, activities,  etc. These textual descriptions can directly guide video generation or filming.
\item \textbf{Auditory Track} specifies the voiceover text and its delivery style. This machine-readable format allows direct application of the generated script in downstream video production. 
\end{itemize}

The MCSC task requires models to jointly perform multimodal comprehension over long and noisy contexts; construct coherent narratives by selecting and creating visual shots; and generate fluent, contextually aligned narrations. It poses a comprehensive challenge for modern MLLMs.

\subsection{Task Evaluation}
\label{sec:eval}

Evaluating video scripts in MCSC requires multi-dimensional assessment, capturing aspects such as visual-audio relevance, narrative coherence, and overall attractiveness. Traditional NLP metrics such as BLEU ~\cite{papineni2002bleu} and ROUGE~\cite{lin2004rouge} fail to reflect these multimodal and creative qualities (see Appendix \ref{sec:bleu}). To address this, we introduce a comprehensive evaluation suite, combining rule-based metrics (Sec. \ref{sec:rule-based}) with multi-dimensional metrics (Sec. \ref{sec:multi_aspect}) that assess aspects like creativity, coherence, and visual-text alignment. 
The multi-dimensional evaluations are performed by fine-tuned expertise MLLMs. As shown in Table~\ref{tab:human_alignment_main} and Appendix \ref{sec:human_eval_details}, these scores correlate well with human judgments, demonstrating their reliability. 

\subsubsection{Automatic Rule-based Metrics}
\label{sec:rule-based}
We define three automatic metrics to quantify common errors in generated scripts:

\noindent\textbf{Error Rate ($Err$)} measures the proportion of distractor materials used in the script.
For example, including shampoo footage in a face cream advertisement. Formally, $Err = \frac{N_{distractor}}{N_{mat}}$, where $N_{distractor}$ counts distractor shots among the total number of material-based shots $N_{mat}$. 

\noindent\textbf{Repetition Rate ($Rep$)} captures redundant usage of video materials, a common issue in less capable MLLMs. Computed as: $Rep = \frac{N_{mat} - N_{unique}}{N_{mat}}$, where $N_{unique}$ is the number of unique material-based shots.

\noindent\textbf{Duration Deviation ($\Delta T$)} evaluates alignment with the user-specified target duration. Defined as: $\Delta T = \frac{|T_{gen} - T_{target}|}{T_{target}}$, where $T_{gen}$ is the generated script’s total duration and $T_{target}$ is the desired duration.

\begin{figure*}[t]
    \centering
    \includegraphics[width=0.95\linewidth]{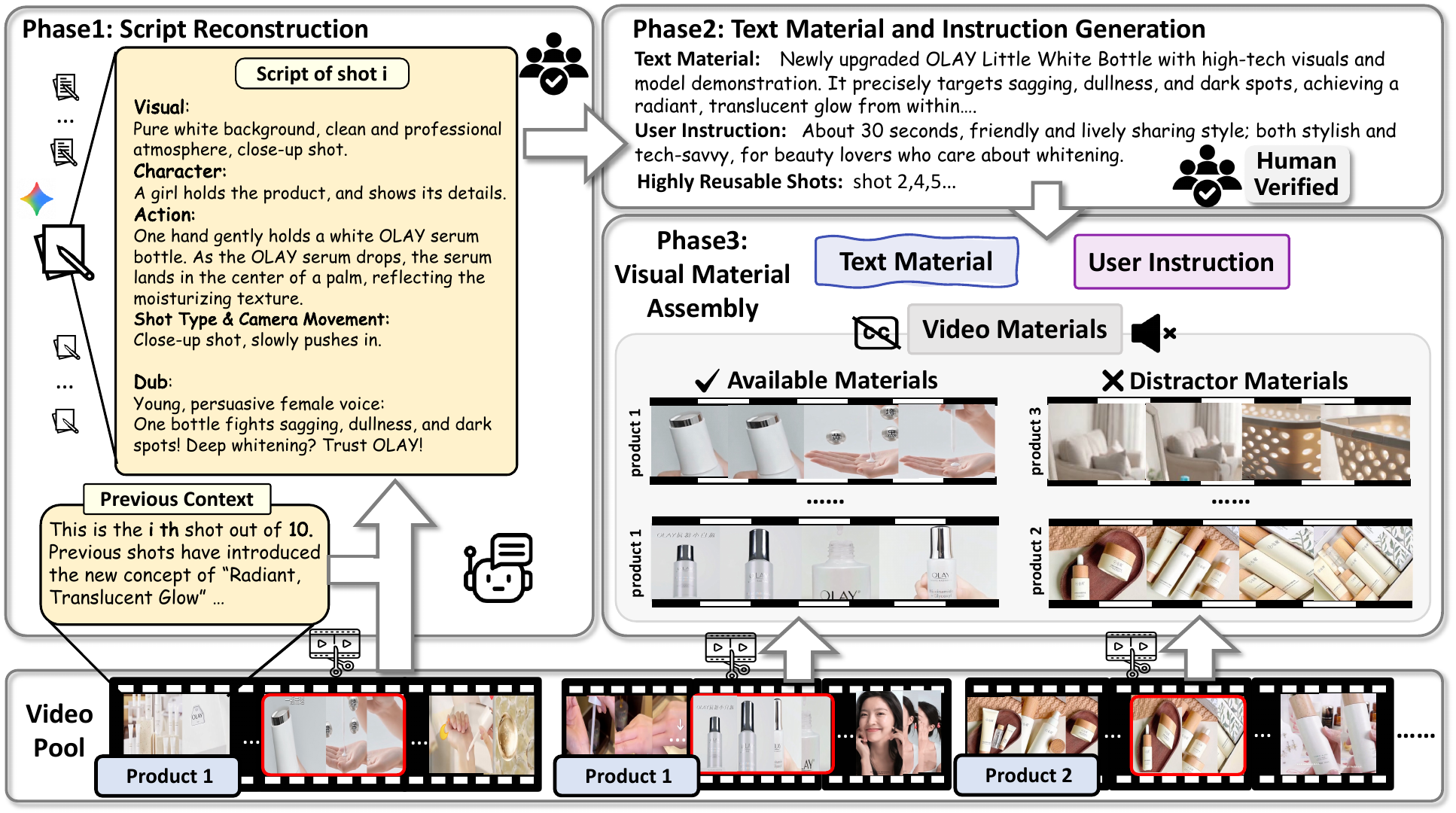}
    \vspace{-8pt}
    \caption{Overview of the MCSC-Bench dataset construction. Video materials are drawn from a large video pool.}
    \label{fig:data_pipeline}
\end{figure*}

\subsubsection{Multi-dimensional Metrics}   
\label{sec:multi_aspect}
To capture semantic, narrative, and creative qualities beyond what rule-based metrics measure, we adopt six multi-dimensional metrics, rated on a 1–5 Likert scale using a fine-tuned open-source MLLM as the judge, which evaluates based on the generated script. Each metric evaluates a distinct aspect of model capability:

\noindent\textbf{Instruction Following (IF):} Fidelity to user instructions, including product features, style, and target audience.

\noindent\textbf{Attractiveness (ATT):} Overall storytelling quality, novelty, engagement, and emotional resonance.

\noindent\textbf{Coherence of Voiceover Text (CVT):} Fluency and logical flow of voiceover text across shots.

\noindent\textbf{Coherence of Visual Sequence (CVS):} Logical ordering and smoothness of the visual sequence, including both material-based and generative shots.

\noindent{\textbf{Voiceover-Visual Alignment (VVA):}} Semantic and temporal alignment between voiceover text and corresponding visuals within each shot.

\noindent\textbf{Necessity of Creation (NC):} Assesses whether the newly planned shots ($S_{\text{new}}$) are necessary, penalizing unnecessary creation when suitable source material already exists.

%% file: tables/tab1_compare.tex
\begin{table*}[t]
\small
\centering
\setlength{\tabcolsep}{1pt}
\begin{tabular}{lcccccc}
\toprule
\textbf{Benchmark} & \textbf{Input} & \textbf{Output} & \textbf{Plot Creation} & \textbf{New Shot Planning} & \textbf{\# Videos} \\ 
\midrule
VScript*~\cite{ji2022vscript} & UI & Script (plot, shots, voiceover) & $\checkmark$ & $\times$ & - \\
Video Storytelling*~\cite{li2019video_storytelling} & V & Story & $\checkmark$ & $\times$ & 105\\
Shot2story*~\cite{han2023shot2story20k} & V & Video Descriptions & $\times$ & $\times$ & 43K \\
Vript*~\cite{yang2024vript} & V & Video Descriptions & $\times$ & $\times$ & 12K  \\
E-SyncVidStory*~\cite{yang2024synchronized} & V \& K & Voiceover & $\times$ & $\times$ & 6K \\
SKYSCRIPT~\cite{tang2024skyscript} & V \& K & Story & $\checkmark$ & $\times$ & 80K \\
Text-to-Edit~\cite{cheng2025text-to-edit} & V \& K \& UI & Voiceover & $\times$ & $\times$ & 100K \\
\midrule
\textbf{MCSC-Bench (Ours)*} & V \& K \& UI & Script (plot, shots, voiceover) & $\checkmark$ & $\checkmark$ & 11K+ \\
\bottomrule
\end{tabular}
\caption{Comparison of our proposed MCSC-Bench with existing benchmarks. In the Input column, V, K, and UI represent video, knowledge, and user instructions, respectively. * denotes open-source benchmarks.}

\label{tab:dataset_comparison}
\end{table*}

%% file: section/4_dataset.tex
\section{MCSC-Bench: Dataset Construction}
\label{sec:data_construct}
To advance research on the MCSC task, we introduce \textbf{MCSC-Bench}, an open-source, large-scale dataset for creative script generation from redundant multimodal materials. As shown in Table~\ref{tab:dataset_comparison}, MCSC-Bench surpasses existing datasets in both task complexity and content richness, requiring models to: process long-context, noisy multimodal inputs; create a coherent plot; and plan new shots. This section describes the data collection, annotation pipeline, and dataset statistics.

\subsection{Data Collection}

To construct a large-scale dataset reflecting real-world video production, we collect professionally produced videos \footnote{High-quality videos with high view counts and purchase rates.} from a popular e-commerce platform~\cite{taobao}, yielding a corpus of 10,726 advertisements spanning 15 product categories (e.g., foods, cosmetics, clothing). To evaluate cross-domain generalization, we further compile a set of 521 general-domain videos (e.g., tutorials, vlogs, global brand ads and short dramas) from public video platforms (YouTube~\cite{youtube} and TikTok~\cite{tiktok}).

\subsection{Dataset Annotation} 
\label{sec:data_pipeline}

We design an automated pipeline to construct both the inputs and ground-truth outputs for the MCSC task. As illustrated in Figure~\ref{fig:data_pipeline}, the pipeline consists of three core phases with rigorous human annotation and revision to ensure quality.

\noindent \textbf{Script Reconstruction.} 
Each video is decomposed into a structured script. We first segment videos into discrete shots using SceneDetect~\cite{scenedetect}. A powerful MLLM (Gemini-2.5-Pro~\cite{comanici2025gemini}) then reconstructs a detailed script for each shot, including a visual description and corresponding voiceover narration. To maintain contextual consistency, the model receives each shot's temporal position and a brief summary of preceding content during reconstruction.

\begin{figure}[t]
    \centering
    \includegraphics[width=0.95\linewidth]{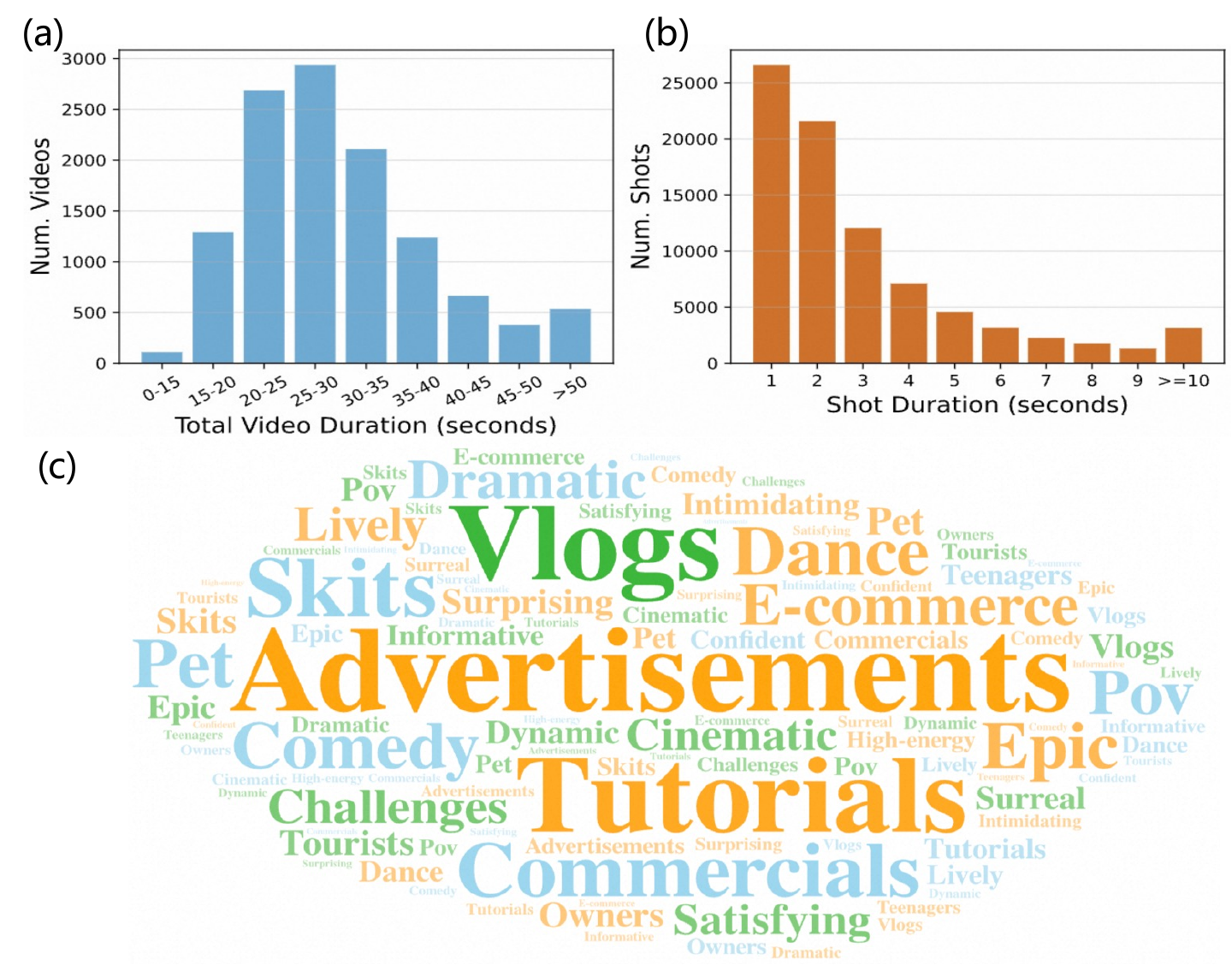}
    \caption{Statistics of MCSC-Bench. (a): Distribution of total video duration. (b): Distribution of shot duration. (c): Word Cloud of Diverse Product Types and Video Types.}
    \label{fig:distribution}
\end{figure}

\noindent \textbf{Text Material and Instruction Generation.} From the reconstructed scripts, Gemini further summarizes the entire script into two textual inputs:
\textbf{text material ($\mathcal{T}$)}, which captures metadata such as core information, keywords or selling points, and \textbf{user instructions ($\mathcal{I}$)}, which specify high-level constraints such as style, target audience, and desired duration.
In addition, the model assigns a \textit{reusability score} to each shot, estimating its potential for reuse across different videos. Shots with high reusability scores are selected as \textit{Relevant Materials} ($\mathcal{V}_a$). Rigorous human annotation and empirical analysis confirm that this process produces high-quality annotations with minimal bias toward the annotating model.

\noindent \textbf{Visual Material Assembly.} To simulate realistic redundant multimodal contexts ($\mathcal{V}$), we construct a visual material pool for each target video that includes two types of clips: (1) \textit{Relevant Materials}, high-reusability shots extracted 
from semantically related videos (e.g., videos sharing the same 
product or subject), and (2)~\textit{Distractor Materials}, randomly sampled shots from 
unrelated videos that serve as noise for material selection. To prevent information leakage, all shots are preprocessed to remove original audio and subtitles using an image editing model \cite{fang2025tbstareditimageeditingpattern}.

\noindent \textbf{Quality Assurance.} To verify quality, 5 professional annotators reviewed 1,500 scripts covering nearly all test data, achieving a 95.7\% production-ready rate; a subsequent re-evaluation by 3 independent annotators yielded 92\% inter-annotator agreement. Besides, script reconstruction primarily extracts objective elements (e.g., ASR transcripts, actions), and evaluated models never receive Gemini-generated scripts as input, eliminating circular evaluation.
A cross-model bias check using Qwen-VL-Max as an alternative annotator further confirms MCSC-Bench does not favor a specific model. Details are in Appendix \ref{sec:anno_details}.

\subsection{Dataset Statistics}
\textbf{MCSC-Bench} contains a total of 11,247 high-quality videos. As shown in Figure~\ref{fig:distribution}, the dataset spans a diverse range of durations, product types, and video types.
On average, each input context contains 7.2 video shots: 5.7 \textit{Relevant Materials} and 1.5 \textit{Distractor Materials}. Textual inputs average 184 words for text material $\mathcal{T}$ and 77 words for user instructions $\mathcal{I}$. These long, noisy, and heterogeneous multimodal inputs form a challenging testbed for multimodal comprehension, content selection, and creative structural script generation.

%% file: section/5_1_experiment.tex
\section{Experiments}

\input{tables/tab2_main_results}

\input{tables/tab4_ood_result}

\subsection{Experimental Settings}
\label{sec:implementation}
We evaluate a wide range of state-of-the-art MLLMs on MCSC-Bench. Each model receives Relevant and Distractor Materials (sampled at 1 fps, audio and subtitles removed), text materials ($\mathcal{T}$), user instructions ($\mathcal{I}$), and is required to generate a structured JSON-formatted script following the schema defined in Section \ref{sec:task_design}.

\paragraph{Evaluated Models.} We evaluate several MLLMs in two categories. Open-source models include Qwen2.5-VL-Instruct (7B, 32B, 72B; abbreviated as Qwen2.5-VL-xB)~\cite{bai2025qwen2}, Qwen3-VL-Instruct (4B, 8B; abbreviated as Qwen3-VL-xB)~\cite{bai2025qwen3}, and InternVL3 (8B, 14B, 38B)~\cite{zhu2025internvl3}. Closed-source models include Qwen-VL-Max~\cite{bai2023qwenmax}, GPT-4.1~\cite{openai2024gpt41}, and Gemini-2.5-Pro~\cite{comanici2025gemini}.

\paragraph{MLLM-based Evaluation.} As described in Section \ref{sec:multi_aspect}, we conduct a systematic evaluation across six dimensions. 
To ensure reproducible evaluation without reliance on proprietary APIs, we train an evaluator model by fine-tuning Qwen2.5-VL-7B on annotations generated by GPT-5~\cite{openai2025gpt5}, which scores scripts generated by diverse models in advance. As shown in Table~\ref{tab:human_alignment_main}, the evaluator achieves Spearman's $\rho > 0.56$ across all six dimensions, consistent with established MLLM-based evaluation methods~\cite{liu2023geval,he2024videoscore}. These results verify that our specialized evaluator serves as an approximate and accessible proxy for high-quality assessment in the MCSC task. Details are in Appendix \ref{sec:human_eval_details}.

\input{tables/tab3_human_main}

\paragraph{Training and Test Setup.} The training set consists of 9,752 in-domain MCSC samples. We fine-tune Qwen3-VL-4B and 8B (\textit{MCSC-4B, MCSC-8B}) using full-parameter fine-tuning for 3 epochs. We further optimize MCSC-8B by applying Group Relative Policy Optimization (GRPO)~\cite{shao2024deepseekmath}, resulting in \textit{MCSC-8B-RL}. Specifically, we use the trained evaluator to score generated scripts on six multidimensional metrics, and use the average normalized score as the reward signal. To mitigate reward hacking, we additionally validate the RL-trained model with an independent judge (details are provided in Appendix \ref{sec:train}). We evaluate the evaluated models and our trained models on the in-domain test set (974 samples), and assess the cross-domain generalization on the out-of-domain test set (521 samples).

%% file: tables/tab2_main_results.tex
\begin{table*}[t]
\centering
\small
\sisetup{table-format=1.2, table-number-alignment = center, tight-spacing=true}
\setlength{\tabcolsep}{4pt}
\begin{tabular}{@{} l 
                   rrr
                  @{\hspace{3.5\tabcolsep}}
                   ccccccc @{}}
\toprule
\multirow{2}{*}{\textbf{Model}} & \multicolumn{3}{c}{\textbf{Automatic Metrics (\%)}} & \multicolumn{7}{c}{\textbf{Multi-dimensional Metrics}} \\
\cmidrule(lr{1em}){2-4} \cmidrule(lr){5-11} 
& {\textbf{$Err$\,↓}} & {\textbf{$Rep$\,↓}} & {\textbf{$\Delta T$\,↓}} & {\textbf{IF\,↑}} & {\textbf{ATT\,↑}} & {\textbf{CVT\,↑}} & {\textbf{CVS\,↑}} & {\textbf{VVA\,↑}} & {\textbf{NC\,↑}} & {\textbf{Avg\,↑}} \\
\midrule
\multicolumn{11}{l}{ \cellcolor{blue!8}\textbf{In-Domain Test} } \\
\midrule
\multicolumn{11}{c}{\cellcolor{lightgray!50}{\textit{Open-source Models}}} \\
Qwen2.5-VL-7B$^\dagger$~\citep{bai2025qwen2}    & 20 & 17 & 180 & 1.35 & 1.88 & 1.86 & 1.73 & 1.50 & 1.51 & 1.64 \\
Qwen2.5-VL-32B~\citep{bai2025qwen2}              & 9  & 12 & 12  & 3.13 & 3.32 & 3.47 & 2.99 & 2.67 & 3.53 & 3.19 \\
Qwen2.5-VL-72B$^\ddagger$~\citep{bai2025qwen2}   & 8  & 4  & 10  & 3.25 & 3.44 & 3.58 & 3.11 & 2.75 & 3.68 & 3.30 \\
InternVL3-8B~\cite{zhu2025internvl3}              & 17 & 32 & 128 & 1.53 & 2.19 & 2.10 & 1.88 & 1.70 & 1.89 & 1.88 \\
InternVL3-14B~\cite{zhu2025internvl3}             & 19 & 21 & 20  & 2.27 & 2.61 & 2.77 & 2.31 & 2.11 & 2.36 & 2.40 \\
InternVL3-38B~\cite{zhu2025internvl3}             & 18 & 5  & 22  & 2.49 & 2.88 & 2.89 & 2.46 & 2.18 & 3.15 & 2.68 \\
Qwen3-VL-4B~\citep{bai2025qwen3}                 & 11 & 2  & 23  & 2.92 & 3.29 & 3.49 & 2.92 & 2.63 & 2.96 & 3.04 \\
Qwen3-VL-8B$^\mathsection$~\citep{bai2025qwen3}  & 12 & 10 & 21  & 3.04 & 3.51 & 3.65 & 2.96 & 2.76 & 3.31 & 3.21 \\
\hdashline
\textit{MCSC-4B}*  & \textbf{0} & \textbf{0} & 7 & 3.43 & 3.65 & 3.60 & 3.14 & 3.05 & 3.89 & 3.46 \\
\textit{MCSC-8B}*  & \textbf{0} & \textbf{0} & 5 & 3.55 & 3.70 & 3.65 & 3.21 & 3.12 & 3.98 & 3.54 \\
\textit{MCSC-8B-RL}*  & 2 & \textbf{0} & 3 & \textbf{4.19} & 4.00 & \textbf{4.03} & \textbf{3.86} & \textbf{3.88} & 4.82 & \textbf{4.13} \\
\midrule
\multicolumn{11}{c}{\cellcolor{lightgray!50}{\textit{Closed-source Models}}} \\
Qwen-VL-Max~\cite{bai2023qwenmax}        & 9 & 10 & 9 & 3.47 & 3.77 & 3.89 & 3.21 & 3.05 & 4.25 & 3.61 \\
GPT-4.1~\cite{openai2024gpt41}           & 3 & 11 & 9 & 3.79 & 3.96 & 4.00 & 3.60 & 3.35 & 4.67 & 3.89 \\
Gemini-2.5-Pro~\cite{comanici2025gemini} & 2 & 10 & \textbf{1} & 4.02 & 4.00 & 4.00 & 3.74 & 3.69 & \textbf{4.84} & 4.05 \\
\midrule
\multicolumn{11}{c}{\cellcolor{lightgray!50}{\textit{Multi-Agent Workflow}}} \\
Qwen2.5-VL-7B$^\dagger$ + Multi-Agent    & 9 & 9 & 32 & 2.21 & 2.92 & 2.82 & 2.67 & 2.34 & 2.92 & 2.65 \\
Qwen2.5-VL-72B$^\ddagger$ + Multi-Agent  & 4 & 1 & 2  & 3.43 & 3.58 & 3.64 & 3.32 & 3.03 & 3.99 & 3.50 \\
Qwen3-VL-8B$^\mathsection$ + Multi-Agent & 1 & 1 & 2  & 3.94 & \textbf{4.01} & 4.00 & 3.76 & 3.70 & 4.82 & 4.04 \\
\bottomrule
\end{tabular}
\caption{
Performance comparison on MCSC-Bench. * denotes fine-tuned models. \textbf{Bold} indicates the best performance. Matching symbols ($\dagger, \ddagger, \mathsection$) link base models to their Multi-Agent counterparts. Avg means the average score of the 6 multi-dimensional metrics.}

\label{tab:main_results}
\end{table*}

%% file: tables/tab4_ood_result.tex
\begin{table*}[t]
\centering
\small 
\sisetup{table-format=1.2, table-number-alignment = center, tight-spacing=true}
\setlength{\tabcolsep}{4pt}
\begin{tabular}{@{} l 
                   rrr
                  @{\hspace{3\tabcolsep}}
                   ccccccc @{}}
\toprule
\multirow{2}{*}{\textbf{Model}} & \multicolumn{3}{c}{\textbf{Automatic Metrics (\%)}} & \multicolumn{7}{c}{\textbf{Multi-dimensional Metrics}} \\
\cmidrule(lr{1em}){2-4} \cmidrule(lr){5-11} 
& {\textbf{$Err$}↓} & {\textbf{$Rep$}↓} & {\textbf{$\Delta T$ }↓} & {\textbf{IF}↑} & {\textbf{ATT}↑} & {\textbf{CVT}↑} & {\textbf{CVS}↑} & {\textbf{VVA}↑} & {\textbf{NC}↑} & {\textbf{Avg}↑} \\
\midrule
\multicolumn{11}{l}{\cellcolor{orange!10}\textbf{ Out-Of-Domain Test }} \\
\midrule
\multicolumn{11}{c}{\cellcolor{lightgray!50}{\textit{Open-source Models}}} \\
Qwen2.5-VL-7B$^\dagger$~\citep{bai2025qwen2}    & 25 & 9  & 121 & 1.92 & 2.57 & 2.01 & 2.59 & 1.87 & 2.43 & 2.23 \\
Qwen2.5-VL-32B~\citep{bai2025qwen2}   & 17 & 7  & 12  & 3.11 & 3.48 & 2.89 & 3.48 & 2.85 & 3.72 & 3.26 \\
Qwen2.5-VL-72B$^\ddagger$~\citep{bai2025qwen2}   & 15 & 6  & 9   & 3.20 & 3.47 & 2.97 & 3.58 & 2.84 & 3.84 & 3.32 \\
InternVL3-8B\cite{zhu2025internvl3}   & 9  & 17 & 45  & 2.39 & 2.96 & 2.36 & 2.99 & 2.30 & 3.00 & 2.67 \\
InternVL3-14B\cite{zhu2025internvl3}  & 22 & 19 & 31  & 2.62 & 3.05 & 2.39 & 3.11 & 2.45 & 2.87 & 2.75 \\
InternVL3-38B\cite{zhu2025internvl3}  & 21 & 8  & 21  & 2.82 & 3.19 & 2.57 & 3.28 & 2.60 & 3.20 & 2.94 \\
Qwen3-VL-4B~\citep{bai2025qwen3}      & 10 & 4  & 27  & 2.94 & 3.22 & 2.77 & 3.39 & 2.67 & 3.12 & 3.02 \\
Qwen3-VL-8B$^\mathsection$~\citep{bai2025qwen3}      & 15 & 4  & 21  & 3.06 & 3.37 & 2.80 & 3.56 & 2.79 & 3.24 & 3.14 \\
\hdashline
\textit{MCSC-4B}*  & \textbf{1} & \textbf{0} & 6 & 3.23 & 3.47 & 3.03 & 3.21 & 2.80 & 3.87 & 3.27 \\
\textit{MCSC-8B}*  & 2 & \textbf{0} & 5 & 3.28 & 3.49 & 3.05 & 3.33 & 2.79 & 3.98 & 3.32 \\

\textit{MCSC-8B-RL}*  & 2 & 1 & 4 & \textbf{3.75} & \textbf{3.92} & \textbf{3.90} & 3.48 & 3.48 & \textbf{4.62} & \textbf{3.86} \\

\midrule
\multicolumn{11}{c}{\cellcolor{lightgray!50}{\textit{Closed-source Models}}} \\
Qwen-VL-Max\cite{bai2023qwenmax}        & 16 & 7  & 11 & 3.33 & 3.67 & 3.00 & 3.77 & 3.07 & 4.00 & 3.47 \\
GPT-4.1\cite{openai2024gpt41}           & 11 & 10 & 10 & 3.59 & 3.88 & 3.45 & \textbf{3.88} & 3.39 & 4.55 & 3.79 \\
Gemini-2.5-Pro\cite{comanici2025gemini} & 3  & 9  & \textbf{0}  & 3.73 & 3.81 & 3.50 & 3.70 & \textbf{3.51} & 4.46 & 3.78 \\
\midrule
\multicolumn{11}{c}{\cellcolor{lightgray!50}{\textit{Multi-Agent Workflow}}} \\
Qwen2.5-VL-7B$^\dagger$ + Multi-Agent    & 10 & 12 & 31 & 2.44 & 3.02 & 2.52 & 3.07 & 2.45 & 3.01 & 2.75 \\
Qwen2.5-VL-72B$^\ddagger$ + Multi-Agent & 2  & 4  & 1  & 3.23 & 3.47 & 3.04 & 3.49 & 2.89 & 3.52 & 3.27 \\
Qwen3-VL-8B$^\mathsection$ + Multi-Agent    & 3  & 6  & 4  & 3.64 & 3.88 & 3.45 & 3.87 & 3.45 & 4.48 & 3.79 \\
\bottomrule
\end{tabular}
\caption{
Performance comparison on the Out-Of-Domain test set.
}
\label{tab:ood_result}
\end{table*}

%% file: tables/tab3_human_main.tex
\begin{table}[h]
\centering 
\small
\begin{tabular}{l ccc}
\toprule
\multirow{2}{*}{\textbf{Metric}} & \multicolumn{3}{c}{\textbf{Ours vs. Human}} \\
\cmidrule(lr){2-4}
& \textbf{Spearman} ($\rho$) & \textbf{Kendall} ($\tau$) & \textbf{Pearson} ($r$) \\
\midrule
IF  & 0.6200* & 0.5350* & 0.6379* \\
ATT & 0.6162* & 0.5281* & 0.6079* \\
CVT & 0.5604* & 0.4823* & 0.5540* \\
CVS & 0.5601* & 0.4818* & 0.5794* \\
VVA & 0.5752* & 0.4919* & 0.5807* \\
NC  & 0.5654* & 0.4819* & 0.5664* \\
\bottomrule
\end{tabular}
\caption{Alignment analysis between our Evaluator Model and human judgments. * denotes p-value < 0.001, indicating statistical significance.}
\label{tab:human_alignment_main}
\end{table}

%% file: section/5_2_capacity.tex
\begin{figure}[h]
    \centering
    \includegraphics[width=1.0\linewidth]{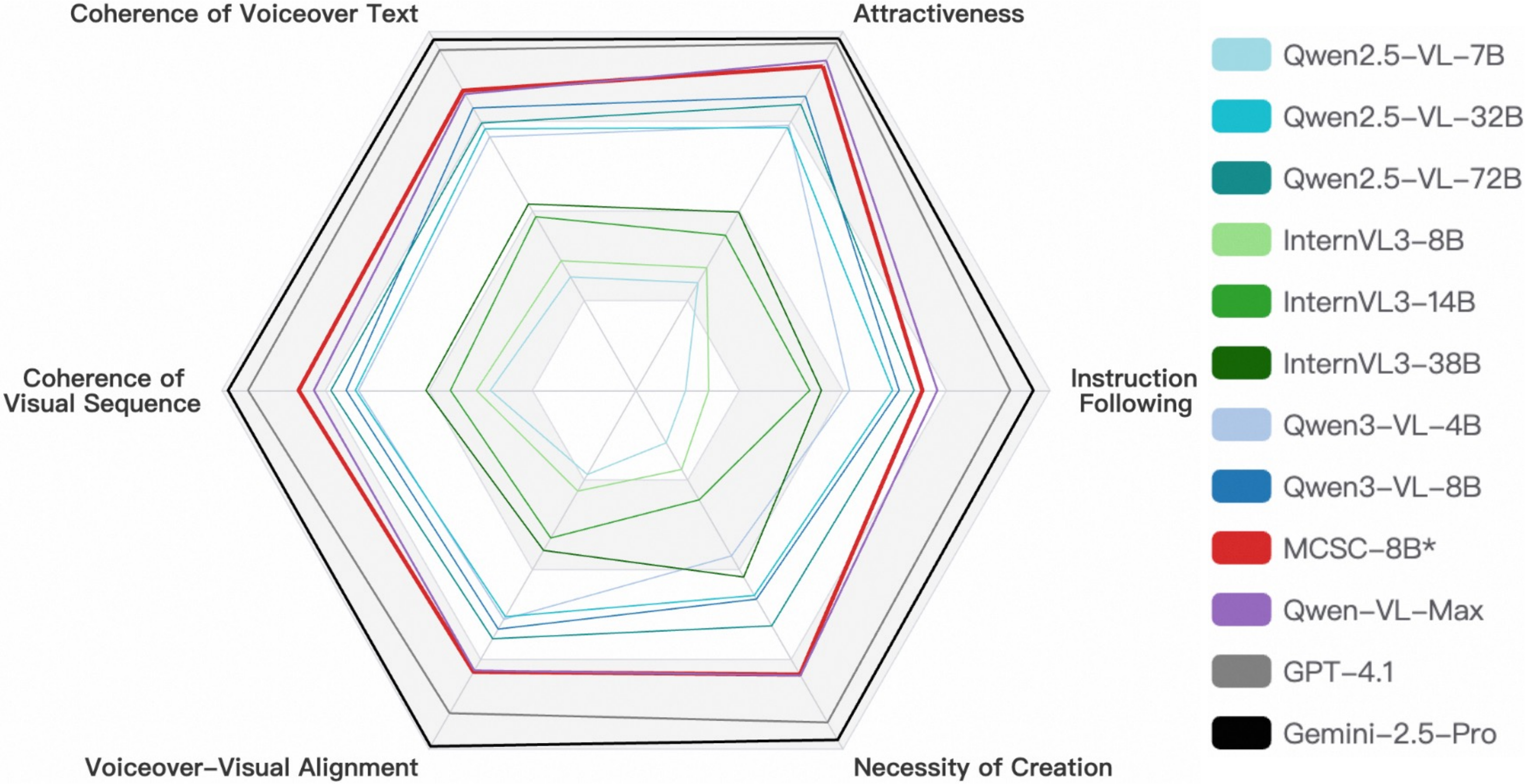}
    \caption{Multi-dimensional evaluation on MCSC-Bench (rescaled by maximum and minimum for better visualization) shows clear performance ladder. }
    \label{fig:radar}
\end{figure}

\subsection{Results and Analysis}

\label{model_capacity}
We benchmark all models on the In-Domain test set (Table~\ref{tab:main_results}) and Out-of-Domain test set (Table~\ref{tab:ood_result}). Results reveal several observations as follows:

\noindent \textbf{Performance Scales with Model Size.} 
Across nearly all metrics, larger models consistently outperform smaller ones. 
This trend demonstrates that MCSC-Bench effectively differentiates model capabilities and poses non-trivial demands on multimodal reasoning and creative generation. 

\noindent \textbf{Global Planning Remains Challenging.} Metrics requiring long-horizon reasoning, particularly \textit{Duration Deviation} ($\Delta T$) and \textit{Instruction Following} (IF), exhibit larger performance gaps, highlighting that global temporal and structural planning remains a key bottleneck.

\noindent \textbf{In-Domain (ID) vs. Out-of-Domain (OOD) Test.}
For most metrics, the benchmarking results on the OOD test set are consistent with our findings on the ID set. Notably, we observe that on the OOD set, \textit{Error Rate (Err)} increases and \textit{Coherence of Voiceover Text (CVT)} drops across most models. This might be attributed to two characteristics of general-domain videos: the presence of semantically similar materials that are harder to distinguish; and the use of diverse narration styles. These factors make both accurate material selection and coherent cross-shot voiceover generation  more challenging.

\noindent \textbf{Training Effectiveness and OOD Generalization.}
Supervised fine-tuning substantially improves performance (MCSC-4B/8B vs. Qwen3-VL-4B/8B), particularly on rule-based metrics that require strict structural control. RL further boosts multi-dimensional scores (MCSC-8B-RL vs. MCSC-8B). Notably, on the general-domain OOD subset, trained models exhibit strong generalization and maintain competitive performance on most metrics, demonstrating that the internal logic of constraint satisfaction and creative generation transfers effectively across domains. These results suggest that MCSC-Bench provides effective and generalizable supervision for the MCSC task.

%% file: section/5_3_multi_agent.tex
\subsection{Multi-Agent Workflow}
\label{sec:multi-agent}

To reduce the cognitive burden induced by entangled subtasks in long multimodal contexts, we introduce a three-stage multi-agent workflow that decomposes MCSC into three stages: analysis, planning, and generation. Details are at Appendix \ref{sec:multi_agent_appendix}.

\noindent \textbf{Strategy.}
The workflow consists of three specialized agents.
The \textbf{Analyzer Agent} performs content analysis by processing multimodal inputs and filtering redundant materials.
The \textbf{Planner Agent} conducts {narrative planning} based on the selected shots, textual material ($\mathcal{T}$), and user instructions ($\mathcal{I}$),  determining shot order and identifying where newly planned shots ($S_{\text{new}}$) are required.
The \textbf{Writer Agent} produces the structured script by converting the plan into JSON-formatted output with aligned visual descriptions and narration. 
The Analyzer outputs a material analysis report for the downstream Planner and Writer, while the Planner outputs a material plan consumed by the Writer. 

\noindent \textbf{Analysis.}
As shown in Table~\ref{tab:main_results} and Table~\ref{tab:ood_result}, the proposed multi-agent workflow consistently improves performance over the end-to-end setting for most models. Decomposing the task mitigates cognitive overload, as evidenced by reductions in $Err$ and consistent gains in reasoning-related metrics such as IF. This confirms that MCSC poses significant cognitive demands on models and can benefit from explicit task decomposition.

%% file: section/5_4_ablation.tex
\subsection{Long-Context Stress Test}
\begin{figure}[t]
    \centering
    \includegraphics[width=\linewidth]{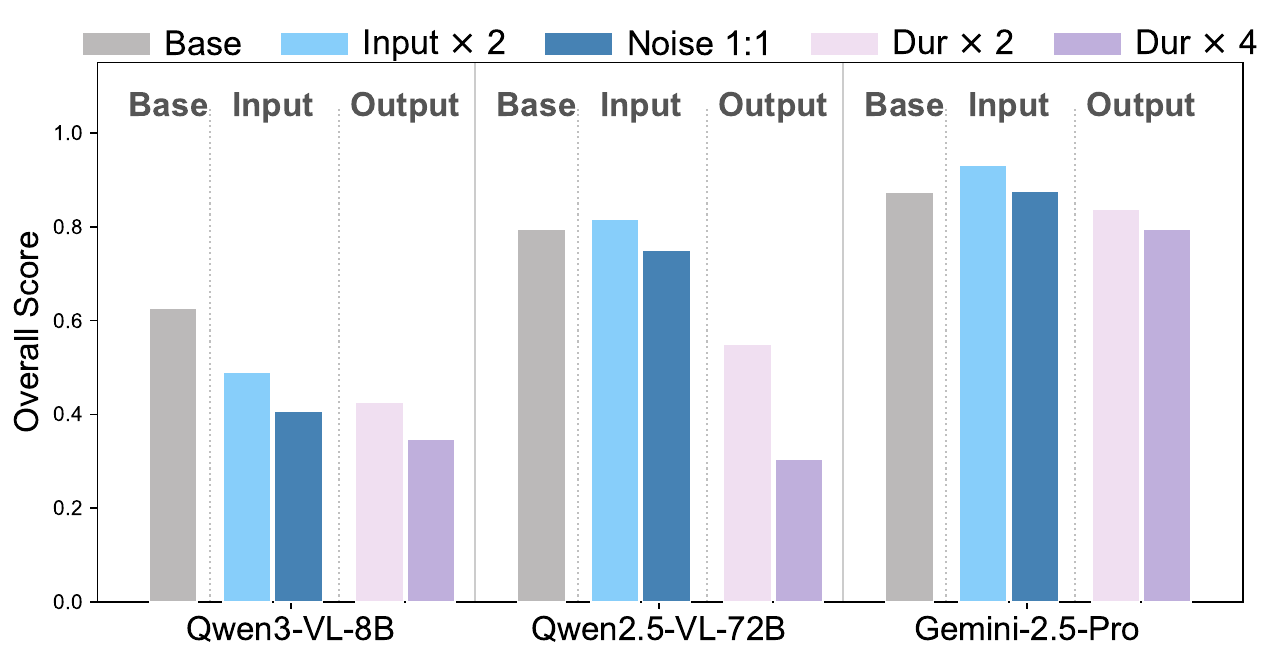}
 \vspace{-8pt}
    \caption{Ablation on Long-Context Stress Test. Base indicates the original input and output setting.}
    \label{fig:ablation_context}
\end{figure}

To examine model robustness under flexible demands, we conduct a comprehensive long-context stress test from both input and output perspectives.
On the Input side, we consider two settings: (1) \textit{Input $\times$ 2}, which increases the average number of shots to 12.43 while maintaining the 4:1 Available-to-Distractor ratio; and (2) \textit{Noise 1:1}, which increases distractor materials to match the number of available materials. 
On the Output side, models are required to produce scripts with Duration $\times$ 2 and Duration $\times$ 4 relative to the target length.
To provide a holistic assessment and discourage degenerate strategies (e.g., trivially short outputs yielding low error rates), we define an Overall Score:
\begin{equation}
    \text{Overall} = (1 - Err) \times (1 - Rep) \times \frac{1}{1 + \Delta T}
\end{equation}
which jointly penalizes material misuse, repetition, and duration deviation. Here, we utilize the continuous penalty term $\frac{1}{1 + \Delta T}$ to prevent the factor from collapsing to zero when $\Delta T$ fluctuates significantly, while $Err$ and $Rep$ are guaranteed to remain positive by their respective definitions and thus are not subject to this issue.

\noindent\textbf{Analysis.} As shown in Figure~\ref{fig:ablation_context}, performance decreases under most stress settings. 
Qwen3-VL-8B exhibits notable sensitivity to both input noise and output length.
Qwen2.5-VL-72B is relatively robust to increased input noise but degrades substantially when longer outputs are required. 
In contrast, Gemini-2.5-Pro shows more stable performance across all dimensions. 
Overall, the results indicate that sustaining effective material selection and planning over extended input and output horizons remains challenging for current MLLMs.

%% file: section/5_5_generation.tex
\subsection{Impact on Video Generation}
\label{sec:video_gen}

While the preceding metrics evaluate script quality at the text level, they do not directly reflect downstream video generation usability. We therefore evaluate whether MCSC scripts can effectively guide video generation models to produce higher-quality videos.

\paragraph{Settings.}
We sample 120 pairs of test scripts, where each pair consists of scripts generated for the same test instance by Qwen3-VL-8B and Gemini-2.5-Pro.
For each script, we compare two video generation strategies:
(1) \textbf{Script-Driven Generation (Ours):} material-based shots ($S_{\text{mat}}$) are taken from original materials, while newly planned shots ($S_{\text{new}}$) are synthesized using \texttt{Wan2.5-T2V}~\citep{wan2025wan} based on script descriptions; (2) \textbf{Instruction-Driven Extension (Baseline):} generative shots are produced using \texttt{Wan2.5-I2V}~\citep{wan2025wan}, conditioned on the last frame (or product image) of the previous shot, text material, and instruction, without explicit narrative planning. We further compare scripts generated by the two models.
We conduct pairwise comparisons using GPT-5~\citep{openai2025gpt5}. Videos are evaluated along three dimensions: Narrative Coherence, Visual Quality, and Overall Appeal. For each pair, GPT-5 selects a winner per dimension. Human evaluation on a subset of 30 pairs in each setting shows strong agreement with GPT-5.

\begin{figure}[t]
\centering
\includegraphics[width=1.0\linewidth]{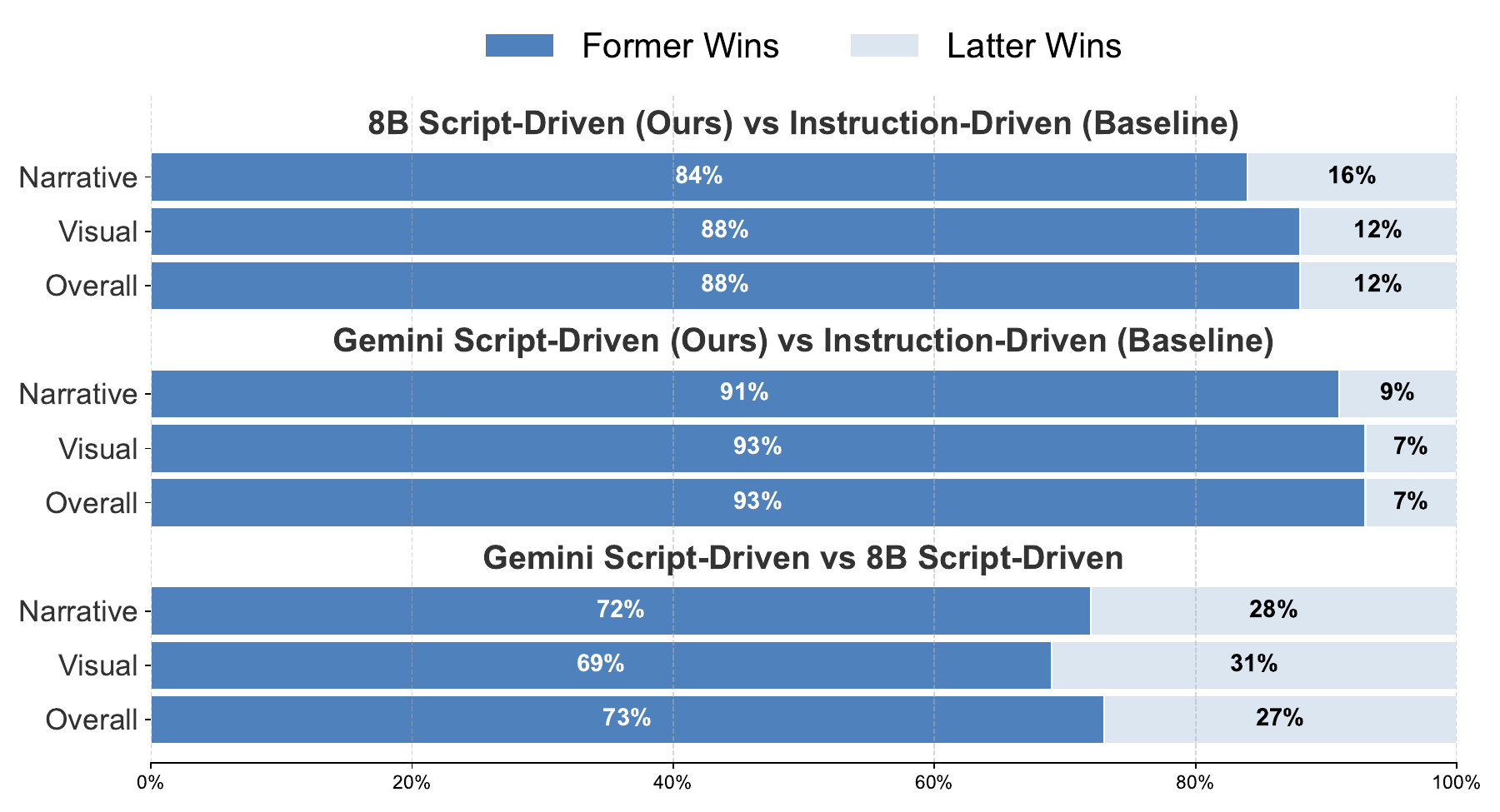}
\caption{Pairwise win rates (\%) for downstream video generation. All differences are statistically significant ($p<0.05$) under two-sided binomial test. }
\label{fig:video_win_rate}
\end{figure}

\begin{figure}[t]
\centering
\includegraphics[width=1.0\linewidth]{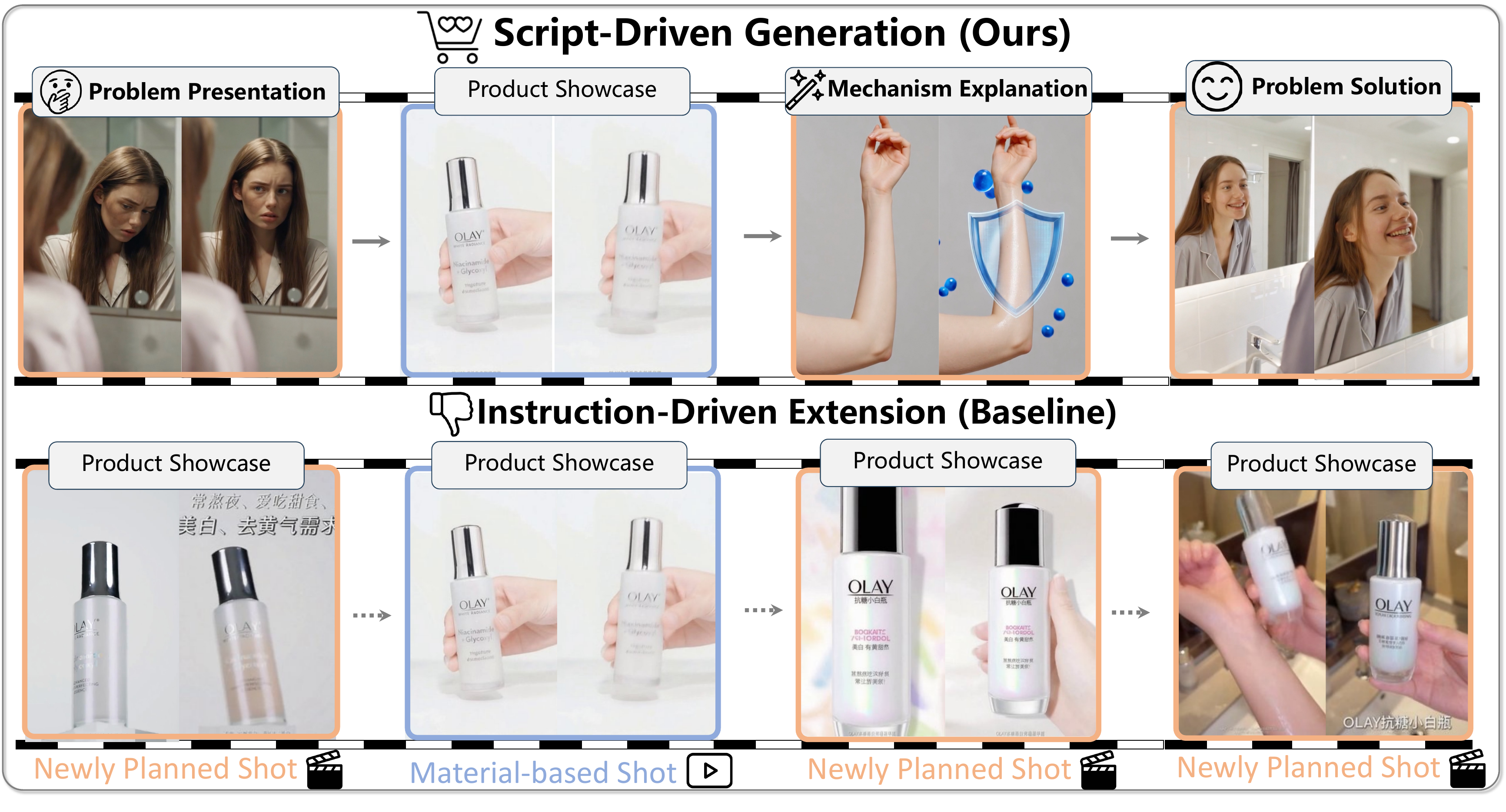}
\vspace{-10pt}
\caption{
Script-Driven (Ours) and Instruction-Driven (Baseline) Video Generation Comparison.}
\label{fig:video_case}
\end{figure}

\paragraph{Results and Analysis.}
Figure~\ref{fig:video_win_rate} shows that Script-Driven Generation consistently outperforms the Instruction-Driven baseline across all dimensions, with overall win rates above 88\% for both models. Qualitative analysis (Figure~\ref{fig:video_case}) indicates that scripts support more coherent narratives and help reduce repetitive showcase shots. In addition, videos generated from Gemini-2.5-Pro scripts outperform those from Qwen3-VL-8B, mirroring trends observed in text-based script evaluation (Table \ref{tab:main_results}). 
These results suggest that MCSC-Bench evaluation correlates with downstream video generation quality. More details are in Appendix \ref{sec:appendix_video_gen}.

%% file: section/2_related_work.tex
\section{Related Works} 

\noindent \textbf{Script Generation.}
Script generation has evolved from text-only inputs \cite{ji2022vscript,chen2024hollmwood,he2024kubrick} to the multi-modal domain\cite{li2019video_storytelling,mahon2024screenwriter,dai2024mmrole,lin2024improving,pennec2025integrating,wu2025automated,chen2025mv-crafter}. Unlike simple video captioning, video-to-script tasks aim to produce structured, production-oriented content. Benchmarks like Vript \cite{yang2024vript}, Shot2Story \cite{han2023shot2story20k}, and SkyScript \cite{tang2024skyscript} have propelled this area by providing large-scale datasets for generating scripts from video. Other related works explore synchronized narration \cite{yang2024synchronized} or text-based editing \cite{cheng2025text-to-edit}.
While these methods have made valuable contributions to multi-modal video script generation, they often exhibit limitations when faced with the real-world challenge of script creation from long, redundant contexts. Our proposed task aims to address this specific gap.

\noindent \textbf{Multi-modal Large Language Models.}
Recent advances in Multi-modal Large Language Models (MLLMs) \cite{achiam2023gpt,comanici2025gemini,bai2025qwen2} have significantly improved video understanding capabilities \cite{chen2025lvagent,shu2025video}. Several studies \cite{liu2023visual,maaz2023videochatgpt,weng2024longvlm,zhang2024llava,chen2024sharegpt4video} have further enhanced performance on video instruction-following tasks. Meanwhile, large-scale benchmarks \cite{wang2024lvbench,li2024mvbench,fu2025videomme,tan2025allvb,zhou2025mlvu} are expanding the frontier of long-video comprehension.
However, existing MLLMs still face challenges in reasoning across multiple video segments and in producing coherent, long-form text to narrate the videos in long contexts.

%% file: section/6_conclusion.tex
\section{Conclusion}

In this paper, we introduce Multimodal Context-to-Script Creation (MCSC), a task that formalizes realistic video production workflows for creating executable
scripts from multimodal long contexts. To support this task, we present \textbf{MCSC-Bench}, the first large-scale dataset comprising 11K+ professionally annotated samples, together with a comprehensive evaluation suite. Extensive experiments reveal that current MLLMs struggle with the joint demands of MCSC, particularly in global planning and instruction following. The model fine-tuned on our dataset achieves SOTA performance, and downstream video generation experiments validate the practical utility of the produced scripts. We hope MCSC-Bench fosters further research on controllable  video script creation.

\section*{Ethics Statement}
This research was conducted in strict adherence to the Code of Ethics and Professional Conduct. All data used in this work is derived from publicly available websites and does not contain personally identifiable information or offensive content. All video copyrights belong to the original authors. For human evaluation, the annotators we recruited possess a high level of education. They were fairly compensated for their time and effort in rating the generated scripts according to our multi-dimensional evaluation criteria.

%% file: section/Appendix/anno.tex
\subsection{Annotation Details}
\label{sec:anno_details}
\noindent\textbf{Data Filtering.} We first collected raw video ads from a leading e-commerce platform and applied coarse-grained filters to retain only videos with high view counts, high purchase conversion, and broad coverage across multiple product categories. We then removed approximately 10\% of the remaining videos that were deemed low quality (e.g., lacking voice-over, poor visual appeal, or weak advertising intent). After this two-stage filtering, the final curated dataset contains less than 2\% of the original ad pool, ensuring that both our training and test sets are of consistently high quality.

\noindent\textbf{Fact-Based Objective Annotation}: Script reconstruction relies predominantly on extracting objective elements inherently present in the video (e.g., ASR, character actions, camera movements). Since these elements have definitive ground truths within the visual/audio content and are less prone to hallucination, we used SOTA Gemini-2.5-Pro for information extraction. Importantly, we ensure a strict separation of inputs during benchmark testing: the evaluated models are only provided with text materials and user instructions. They do not receive Gemini-generated scripts as inputs, thereby eliminating any circular loop of evaluating Gemini on its own generated text format.

\noindent\textbf{Human Annotation}: After automated annotation, we employed professional, university-educated annotators to manually review and revise the scripts. Specifically, 5 annotators each checked 300 scripts, totaling 1500 scripts, which covers almost all test data. They focused especially on text material and user instructions, which are less objective than video scene descriptions. Annotation included style consistency, factual accuracy, and video-text alignment, with 95.7\% of the scripts identified as "production-ready" (containing no or only minor bias or errors). 

To further guarantee quality, we recruited 3 new annotators to check 100 annotated scripts from the test set for a Pass/Fail re-evaluation. The agreement rate with the original annotations reached a high IAA of 92.3\%, strictly meeting the quality standards for high-tier datasets.

\subsubsection{Cross-Model Bias Check}
To further ensure our dataset does not introduce biases in favor of Gemini, we use Qwen-VL-Max instead of Gemini as the Data Annotator for 150 random samples from the test set in the same data construction pipeline. We then evaluate models on both data annotators. As shown in Table~\ref{tab:bias_check}, across different annotators, Gemini consistently outperforms Qwen in script generation. The IF (instruction following) score, which could be affected by potential bias, shows no clear difference across annotators. Any slight variation stems from generation variance rather than annotation bias.

\begin{table}[h]
\centering
\small
\setlength{\tabcolsep}{2pt}
\begin{tabular}{ccccc}
\toprule
\textbf{Annotator} & \textbf{Generator} & \textbf{Err}(\%) $\downarrow$ & $\Delta T(\%) \downarrow$  & \textbf{IF} $\uparrow$ \\
\midrule
Qwen & Qwen & 11  & 12 & 3.27 \\
Gemini   & Qwen  & 7 & 10  & 3.49 \\
\midrule
Qwen   & Gemini & 1   & 1   & 4.03 \\
Gemini & Gemini & 1   & 1  & 4.06 \\
\bottomrule
\end{tabular}
\caption{\textbf{Analysis of Annotator Bias.} We compare the performance of Script Generators when the underlying data is annotated by different models--Qwen-VL-Max (abbreviated as Qwen) vs. Gemini-2.5-Pro (Gemini). The consistent superiority of Gemini confirms that the performance is not due to annotator-specific bias.}
\label{tab:bias_check}
\end{table}

\subsubsection{More Definition in Annotation}

\noindent\textbf{Distractors}: To ensure a realistic challenge, $\mathcal{V}_{d}$ encompasses other products or scenes from the other videos, requiring the model to distinguish it, which is measured by $Err$, while redundant usage of materials is separately penalized by the $Rep$. 
This formulation simulates real-world workflows where creators must first filter footage from large asset libraries. 
Our experiments reveal that this is a non-trivial setting, with many models exhibiting $Err > 10\%$. 

\noindent\textbf{SceneDetect}: Configured with ContentDetector threshold=50 and AdaptiveDetector threshold=30.

\noindent\textbf{Reusability Score (0-10)}: Measures narrative independence (e.g., high for product shots, low for lip-sync drama).

%% file: section/Appendix/prompt.tex
\subsection{Main Prompt}
Figure \ref{fig:prompt_compose} illustrates the prompt for benchmarking all models. Figure \ref{fig:prompt_evaluation} illustrates the prompt for evaluating models in MCSC-Bench.
Figure \ref{fig:prompt_script} illustrates the prompt for automatic data construction.

%% file: section/Appendix/train.tex
\subsection{Training Details on SFT and RL}
\label{sec:train}
For supervised fine-tuning, we apply the AdamW optimizer (initial learning rate 1\text{e-}5, cosine schedule).

In MCSC-8B-RL, we further apply reinforcement learning using Group Relative Policy Optimization (GRPO)~\cite{shao2024deepseekmath} to enhance response quality on top of MCSC-8B. We apply our trained evaluator to score generated scripts across the 6 dimensions, and take the average normalized score as the reward signal. 
The RL training is conducted with a learning rate of 1e-6 with AdamW optimizer, an effective batch size of 128, and KL regularization ($\beta=0.01$). The vision encoder is frozen.

\begin{table*}[t]
\centering
\small
{
\begin{tabular}{l l c c c c c c}
\toprule
\textbf{Judge} & \textbf{Model} & \textbf{IF\,↑} & \textbf{ATT\,↑} & \textbf{CVT\,↑} & \textbf{CVS\,↑} & \textbf{VVA\,↑} & \textbf{NC\,↑} \\
\midrule
\multirow{2}{*}{\textit{Evaluator (Reward Model)}}
& MCSC-8B        & 3.55 & 3.70 & 3.65 & 3.21 & 3.12 & 3.98 \\
& MCSC-8B-RL     & 4.19 & 4.00 & 4.03 & 3.86 & 3.88 & 4.82 \\
\midrule
\multirow{2}{*}{\textit{Gemini-2.5-Pro (Held-out)}}
& MCSC-8B        & 4.02 & 3.12 & 3.36 & 2.24 & 2.90 & 3.82 \\
& MCSC-8B-RL     & 4.13 & 3.52 & 4.38 & 2.64 & 3.14 & 4.58 \\
\bottomrule
\end{tabular}
}
\caption{Independent evaluation to verify the absence of reward hacking.}
\label{tab:reward_hacking}
\end{table*}

To rule out the possibility of reward hacking, we additionally evaluate the RL-trained model using an independent evaluator Gemini-2.5-pro ~\cite{comanici2025gemini}. As shown in Table \ref{tab:reward_hacking}, the performance gains of MCSC-8B-RL over MCSC-8B remain consistent under this independent evaluation, confirming that the improvements reflect genuine script quality enhancement rather than overfitting to the reward model's scoring biases.

%% file: section/Appendix/bleu.tex
\subsection{Traditional NLP Metrics}
\label{sec:bleu}
As shown in Table~\ref{tab:nlp}, based on traditional text-overlap metrics, Gemini-2.5-Pro, MCSC-8B*, MCSC-8B-RL demonstrate better performance, indicating a high degree of lexical similarity with the reference scripts. However, the overall low scores across all models underscore the inherent limitations of these evaluation methods for this task. Script quality is determined by creative and structural elements like narrative logic, cinematic language, and coherence, rather than mere word-for-word matching. A well-written but stylistically different script would be unfairly penalized by these metrics, which are incapable of assessing crucial aspects such as structural validity, semantic accuracy, or narrative plausibility.

\begin{table}[h]
\centering
\small
\setlength{\tabcolsep}{1pt}
\begin{tabular}{lccccc}
\toprule
\textbf{Model} & \textbf{BLEU-1} & \textbf{BLEU-2} & \textbf{BLEU-3}& \textbf{ROUGE-L} \\
\midrule
Qwen2.5-VL-7B      & 15.21 & 5.89 &  2.90 & 18.54 \\
Qwen2.5-VL-32B      & 20.27 & 9.41& 4.89 & 20.56 \\
Qwen2.5-VL-72B     & 16.67 & 7.85 & 4.11 & 19.79 \\
InternVL3-8B    & 15.51 & 5.81 &2.71 & 19.81 \\
InternVL3-14B    & 5.20 & 2.29 & 1.22 & 15.58 \\
InternVL3-38B     & 16.65 & 7.79 & 4.23 &21.05 \\
Qwen3-VL-4B    &10.13 & 4.38 & 2.12 & 15.55 \\
Qwen3-VL-8B   & 19.02   & 8.33    &  4.06  &  18.57 \\
\textit{MCSC-4B}*   & 30.27 &   16.42 &   9.68 & 23.87       \\  \textit{MCSC-8B*}    & 31.37   &  17.43  &  10.48  &24.81   \\
\textit{MCSC-8B-RL}    & 30.22   &  17.22  &  11.02  &22.94   \\
\midrule
Qwen-VL-Max       & 20.16 & 9.05 & 4.61 &  20.37 \\
GPT-4.1       & 23.16 & 9.25 & 4.11 & 19.07 \\
Gemini-2.5-Pro & 26.00 & 11.62 & 5.69 &21.24 \\
\bottomrule
\end{tabular}
\caption{Performance comparison of various models on traditional text generation metrics.}
\label{tab:nlp}
\end{table}

%% file: section/Appendix/eval.tex
\subsection{Evaluator Implementation and Validation}
\label{sec:human_eval_details}

\paragraph{Evaluator Training Details.} 
The construction of our Evaluator Model involved a two-step knowledge distillation process. First, we prompted GPT-5 to score model outputs on our test set and provide detailed reasoning for each dimension. Then, we randomly sampled cases to yield a high-quality dataset of scoring samples (reason-score pairs). Subsequently, we fine-tuned Qwen2.5-VL-7B on this dataset using LoRA fine-tuning (learning rate=1e-5, LoRA rank=128, epoch=3) and 16 NVIDIA A100 GPUs, training it to predict the score for one specific metric at a time.

\paragraph{Human Evaluation Setup.}
To rigorously validate the reliability of our fine-tuned Evaluator Model (henceforth, "Ours"), we conducted a comprehensive human evaluation. The study involved 9 graduate-level annotators who were trained on the task's objectives and evaluation criteria. We randomly sampled a total number of 350 scripts generated by the baseline models. To ensure robust inter-annotator agreement (IAA) analysis, a subset of these scripts was cross-annotated by multiple reviewers, finally leading to a total of 500 unique ratings in each dimension.

Our analysis focuses on three key questions: (1) How well does our Evaluator Model align with human judgment? (2) How does this alignment compare to that of the teacher model, GPT-5? (3) How effectively did our model learn the scoring patterns of GPT-5? Detailed results are presented in Table~\ref{tab:full_correlation_appendix}.

\begin{table*}[h]
\centering 
\small
\setlength{\tabcolsep}{4pt}
\begin{tabular}{l ccc ccc c}
\toprule
\multirow{2}{*}{\textbf{Metric}} & \multicolumn{3}{c}{\textbf{GPT-5 vs. Human}} & \multicolumn{3}{c}{\textbf{Ours vs. GPT-5}} & \textbf{Human IAA} \\
\cmidrule(lr){2-4} \cmidrule(lr){5-7} \cmidrule(lr){8-8}
& \textbf{Spearman} ($\rho$) & \textbf{Kendall} ($\tau$) & \textbf{Pearson} ($r$) & \textbf{Spearman} ($\rho$) & \textbf{Kendall} ($\tau$) & \textbf{Pearson} ($r$) & \textbf{Krippendorff's} $\alpha$ \\
\midrule
IF  & 0.6359 & 0.5495 & 0.6509 & 0.8867 & 0.8257 & 0.8914 & 0.690 \\
ATT & 0.6173 & 0.5371 & 0.6146 & 0.8707 & 0.8173 & 0.8564 & 0.721 \\
CVT & 0.5610 & 0.4849 & 0.5756 & 0.8323 & 0.7757 & 0.8480 & 0.701 \\
CVS & 0.5947 & 0.5110 & 0.5992 & 0.8385  & 0.7720 & 0.8250 & 0.679 \\
VVA & 0.5681 & 0.4806 & 0.5641 & 0.8272 & 0.7471 & 0.8093 & 0.690 \\
NC  & 0.5507 & 0.4578 & 0.5626 & 0.8371 & 0.7527 & 0.8557 & 0.688 \\
\bottomrule 
\end{tabular}
\caption{
Additional validation statistics. We compare the correlation of GPT-5 against Human judgment, and our Evaluator against GPT-5. All results are \textbf{statistically significant} ($p < 0.0001$). 
}
\label{tab:full_correlation_appendix}
\end{table*}

\paragraph{Alignment with Human Judgment.} 
We first measured the correlation between our Evaluator Model's scores and the human preference scores. The results show statistically significant positive correlations across all six dimensions, with Spearman's $\rho$ coefficients ranging from 0.56 to 0.62 (in the main paper). These values are highly comparable to the correlation coefficients between GPT-5 and human annotators (Spearman's $\rho$ ranging from 0.55 to 0.64). This crucial finding confirms that our open-source, reproducible evaluator serves as a reliable proxy for human judgment, achieving a level of alignment that is on par with a state-of-the-art proprietary model.

\paragraph{Effectiveness of Knowledge Distillation.} 
To verify that our fine-tuning process successfully distilled the scoring behavior of the teacher model, we computed the correlation between our evaluator's scores and GPT-5's scores on the same set of outputs. The correlation is exceptionally high, demonstrating that our model has effectively learned and replicated the scoring patterns of GPT-5, validating the SFT strategy.

\paragraph{Reliability of Human Ground Truth.} 
Finally, to confirm the quality of the collected human data, we computed Krippendorff’s $\alpha$ to measure inter-annotator agreement. All metrics surpassed the 0.667 threshold, indicating substantial agreement among annotators and affirming the reliability. The statistical significance of all reported correlations ($p < 0.0001$) further solidifies the validity of our findings.

\paragraph{Upper Bound in Evaluation.}
It is important to recognize that evaluation dimensions such as attractiveness are inherently subjective. Even among human annotators, perfect agreement is unattainable. The human–human inter-annotator agreement (IAA) is approximately 0.70. This naturally imposes an upper bound on achievable machine–human alignment. An evaluator reaching correlations around 0.60 is therefore operating relatively close to the consensus limit of human evaluators. We agree, however, that this does not eliminate the reliability ceiling. Evaluation noise remains non-trivial and our evaluator serves as an  approximate and reproducible proxy rather than ground truth.

%% file: section/Appendix/multi_agent.tex
\subsection{Multi-agent Workflow}
\label{sec:multi_agent_appendix}
\paragraph{Cost/Performance Tradeoffs.}
The Multi-Agent workflow achieves a highly favorable balance: it dramatically reduces critical reasoning errors (e.g., $Err$ drops to near 0\%), while increasing inference time and token usage by only 2.3× and 1.8×, respectively.
Most of the computation occurs in the first stage (multi-video understanding), which we optimize with parallel encoding. The subsequent Planner and Writer stages handle only textual representations, keeping their processing lightweight. As a result, total inference time remains under 8 seconds per query, and token usage grows modestly.
Importantly, end-to-end models often fail with long contexts, producing unusable scripts that require multiple regenerations, which can drastically increase practical costs. In contrast, the Multi-Agent workflow produces highly coherent, production-ready scripts in a single pass, making the bounded increase in time and tokens a worthwhile tradeoff for reliable, high-quality outputs.

\paragraph{Ablation on Decomposition Strategies.} 
To investigate whether all three stages are necessary and to compare with alternative reasoning methods, we conduct ablation studies based on Qwen3-VL-8B. We evaluate Chain-of-Thought (CoT) prompting (which forces the model to explicitly reason before generating the script in a single pass) and two 2-stage variants: (1) \textit{Select+Plan $\rightarrow$ Write}, and (2) \textit{Select $\rightarrow$ Plan+Write}. As shown in Table \ref{tab:ablation_agent}, while CoT and 2-stage methods offer improvements over the base end-to-end setting, they lag behind the full 3-stage decomposition. Notably, the ``select-first'' setting (Select $\rightarrow$ Plan+Write) substantially reduces the error rate ($Err$), whereas the ``joint selection-and-planning'' setting (Select+Plan $\rightarrow$ Write) yields better duration control ($\Delta T$) and overall average score. Ultimately, our 3-stage workflow remains the definitive SOTA across all metrics, demonstrating that isolating each cognitive step is crucial for optimal performance.

\begin{table}[ht]
\centering
\small
\setlength{\tabcolsep}{2pt}
\begin{tabular}{lcccc}
\toprule
\textbf{Setting} & \textbf{$Err\downarrow$} & \textbf{$Rep\downarrow$} & \textbf{$\Delta T\downarrow$} & \textbf{Avg $\uparrow$} \\
\midrule
Base Experiment & 12 & 10 & 21 & 3.21 \\
Chain-of-Thought (CoT) & 7 & 4 & 16 & 3.62 \\
2-Stage (Sel+Plan $\rightarrow$ Write) & 5 & 10 & 6 & 3.78 \\
2-Stage (Sel $\rightarrow$ Plan+Write) & 1 & 4 & 18 & 3.73 \\
\midrule
\textbf{Multi-agent (3-Stage)} & \textbf{1} & \textbf{1} & \textbf{2} & \textbf{4.04} \\

\bottomrule
\end{tabular}
\caption{Ablation of decomposition strategies on Qwen3-VL-8B. The 3-stage workflow yields the best overall performance.}
\label{tab:ablation_agent}
\end{table}

\paragraph{Comparison with Retrieval and Existing Systems.} 
Beyond prompt-based decomposition, we explore a Retrieval-Augmented Generation (RAG) baseline using CLIP-ViT-Large to encode all text and video materials, employing cosine similarity to filter relevant clips. The RAG baseline yields a high error rate, with 23\% of the retrieved clips being distractors, compared to only 1\% for our Analyzer Agent. This confirms that simple semantic retrieval cannot replace deep multimodal reasoning in realistic contexts. Furthermore, compared to existing video script generation systems like VC-LLM, which primarily focus on basic temporal concatenation and dubbing, the MCSC task and our proposed workflow explicitly address the curation of distractor materials, fine-grained multimodal constraint satisfaction, and open-ended generative narrative planning. These systems are not directly optimized for the structured reasoning over heterogeneous materials that MCSC demands.

%% file: section/Appendix/case_study.tex
\subsection{Case Studies}
\label{sec:error_analysis}

This section provides case studies to illustrate common errors observed in the generated scripts (From Figure \ref{fig:7b} to \ref{fig:intern}, high-quality content marked in green, while errors marked in red). A review of the outputs confirms that most models can produce structurally valid and readable scripts containing both material-based shots ($S_{mat}$) and newly planned shots ($S_{new}$), and incorporate product information from $\mathcal{T}$ and user instructions from $\mathcal{I}$ into the script. 

However, a closer inspection reveals several deficiencies, which can be categorized into the following error types:

\noindent\textbf{(1) Erroneous Use of Distractors (~40\%):} This is the most frequent failure. Models incorporate irrelevant distractor materials into the script due to limitations in visual comprehension. This severely degrades the narrative's coherence and relevance, as seen in Figures \ref{fig:7b} and \ref{fig:8b}. It indicates that current models lack fine-grained discrimination in long contexts when experiencing cognitive overload.

\noindent\textbf{(2) Lack of Semantic Organization (~20\%):}    Some models fail to perform genuine content filtering and planning. Instead, they tend to process the input video materials sequentially in their provided (shuffled) order before creating new shots(e.g., Figures \ref{fig:7b} and \ref{fig:intern}). This reflects a limitation in high-level planning and reasoning capabilities (particularly in small models).

\noindent\textbf{(3) Content Repetition (~20\%):} A tendency to generate repetitive content appears, such as the redundant use of the same video material, and the duplication of narrative phrases across different shots (e.g., Figure \ref{fig:intern}), showing limitations in long-term memory.

\noindent\textbf{(4) Other Errors (~20\%)}: Mismatches between generated dialogue and visuals, or failure to meet duration constraints.

\paragraph{Difficult Shot Types.}
\textbf{Style/emotion-related shots are more error‑prone}. For instance, the model may generate a quiet bedroom scene that conflicts with the user’s instruction to create a lively atmosphere. Handling this type of shot requires strong high‑level instruction following.

\paragraph{Case Studies on Out-Of-Domain Test.}

As illustrated in Figure \ref{fig:mcsc8b_gen} and Figure \ref{fig:case_gemini_en_ad}, we observe that general videos exhibit richer visual content, more complex dialogue logic, and more complex narrative structures, thereby posing greater challenges to the model. For instance, in Figure \ref{fig:mcsc8b_gen}, MCSC-8B employs visually similar yet contextually irrelevant video materials.

%% file: section/Appendix/video_generation.tex
\subsection{Video Generation Implementation Details}
\label{sec:appendix_video_gen}

\subsubsection{Input Construction}
The total number of shots in the selected script ranges from 6 to 12, with 3 to 8 being newly planned shots.

We employ distinct prompt construction strategies for each generation mode:

\textbf{Script-Driven Generation} For \texttt{Wan2.5-T2V}, we construct the prompt by concatenating fields from the newly planned shot JSON object using the following template:
\begin{quote}
<setting>. <character>. <action>. 
Camera: <camera\_movement> <shot\_type>.
\end{quote}
    
\textbf{Instruction-Only Extension} For \texttt{Wan2.5-I2V}, the prompt is formatted as:
\begin{quote}
Product: <product\_name>. <style>. <audience>.
\end{quote}
This formulation encourages the video generation model to rely on its internal prior and the visual condition (the last frame) rather than explicit narrative guidance.

\subsubsection{Model Selection Rationale}
We utilized the \texttt{Wan2.5} suite for both settings to minimize backbone discrepancies.
\textbf{T2V for Scripts:} The MCSC task requires the "Creation" of new shots that often involve significant scene shifts (e.g., cutting from a product close-up to a family living room) to serve the narrative. T2V models are essential for realizing these discontinuous scene changes described in the script.
\textbf{I2V for Baseline:} The baseline represents a scenario where the model lacks a "plan" for a new scene. In the absence of a script description, the most natural generation strategy is to extend the visual flow of the preceding shot using I2V. While video generation models may exhibit minor consistency issues (e.g., product identity), our evaluation focuses on narrative structure and overall quality. We treat the consistency issues in generation models as a promising and important future direction.

Additionally, this baseline choice is dictated by the limitations of current video generation architectures. Existing SOTA models lack the native capability to reason over long, redundant multimodal contexts (i.e., dozens of raw clips plus detailed textual constraints) and output the long video in a single pass. 
Consequently, they cannot perform global planning or material selection internally. Therefore, the sequential, locally-conditioned extension (via I2V) represents the most reasonable baseline workflow for automated video production.

\subsubsection{Human Evaluation}
To verify the reliability of the GPT-5-based automatic evaluation for video generation, we conducted a rigorous human evaluation. We randomly selected 30 video pairs from each of the three comparison settings, totaling 90 pairs. 
Three trained annotators (all university-educated) with competence in evaluating video quality were recruited to judge the pairs based on Narrative Coherence, Visual Quality, and Overall Appeal, using the same criteria as GPT-5.
We calculated the Agreement Rate, defined as the percentage of instances where the human vote matched the GPT-5 judgment (Win/Loss/Tie).
As shown in Table~\ref{tab:human_video_eval}, The average agreement rates are high across all dimensions. This confirms that GPT-5 serves as an effective proxy for assessing downstream video generation quality.

\begin{table}[t]
\centering
\small
\setlength{\tabcolsep}{4pt}
\begin{tabular}{@{}l ccc@{}}
\toprule
\textbf{Agreement Rates(\%) } & \textbf{Narrative} & \textbf{Visual} & \textbf{Overall} \\
\midrule
Average & 88.9 & 86.7 & 88.9 \\
\bottomrule
\end{tabular}
\caption{Human-Model Agreement Rates (\%) on the downstream video generation task.}
\label{tab:human_video_eval}
\end{table}

\subsubsection{Video Generation Case Study}

\paragraph{Impact of Script.}
Figure~\ref{fig:video_case1} demonstrates the structural superiority of the script-driven approach. The script-driven video (Ours) exhibits a clear narrative arc: it begins by establishing a user need, transitions to the product as a solution, and concludes professionally with a brand slogan and logo. This "Problem-Solution-Branding" structure effectively drives emotional engagement. In contrast, the Instruction-Driven baseline lacks this progression. It tends to generate repetitive product close-ups or generic text overlays, failing to construct a coherent story or shift scenes to show diverse usage scenarios.

\paragraph{Differences in Script Quality.}
Figure~\ref{fig:video_case2} highlights the performance gap between Qwen3-VL-8B and Gemini-2.5-Pro. Gemini-generated scripts result in more attractive visual storytelling. For instance, in the skincare example, Gemini employs a dramatic contrast: transitioning from a happy girl playing with bubbles to a close-up of skin problems. It enhances the need for the product, whereas Qwen produces a more linear and less emotionally hooky script. 
Similarly, in the festive gift example, Gemini showcases advanced pacing by cutting between multiple scenarios to emphasize the product's versatility and festive atmosphere. Qwen's output remains relatively static and occasionally includes unrelated shots, confirming that superior reasoning models produce richer and more attractive visual blueprints.

%% file: section/Appendix/image.tex
\begin{figure*}[t]
    \centering
    \includegraphics[width=\textwidth]{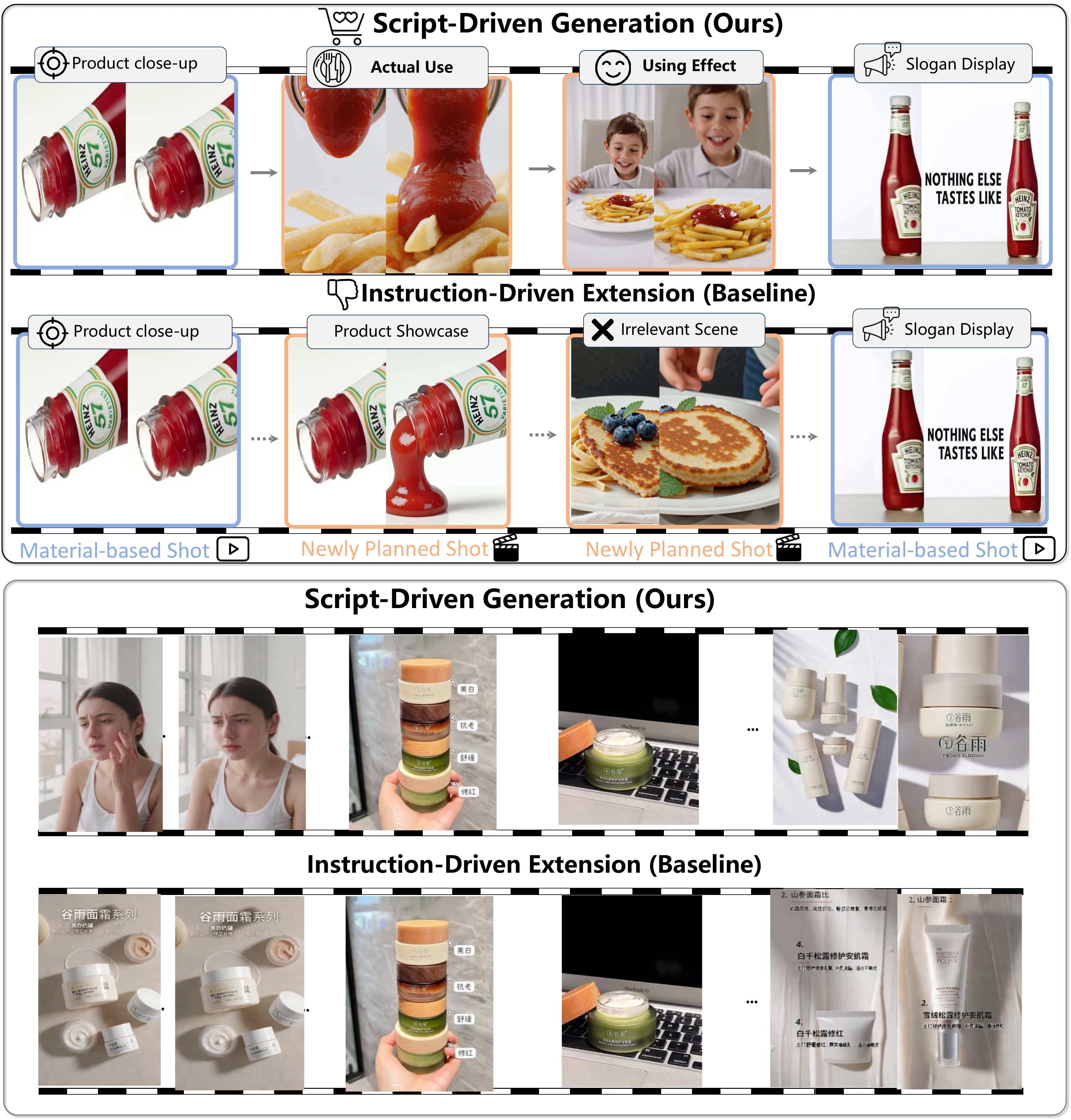}
    \caption{Qualitative comparison between our Script-Driven approach and the Instruction-Driven baseline.}
    \label{fig:video_case1}
\end{figure*}

\begin{figure*}[t]
    \centering
    \includegraphics[width=\textwidth]{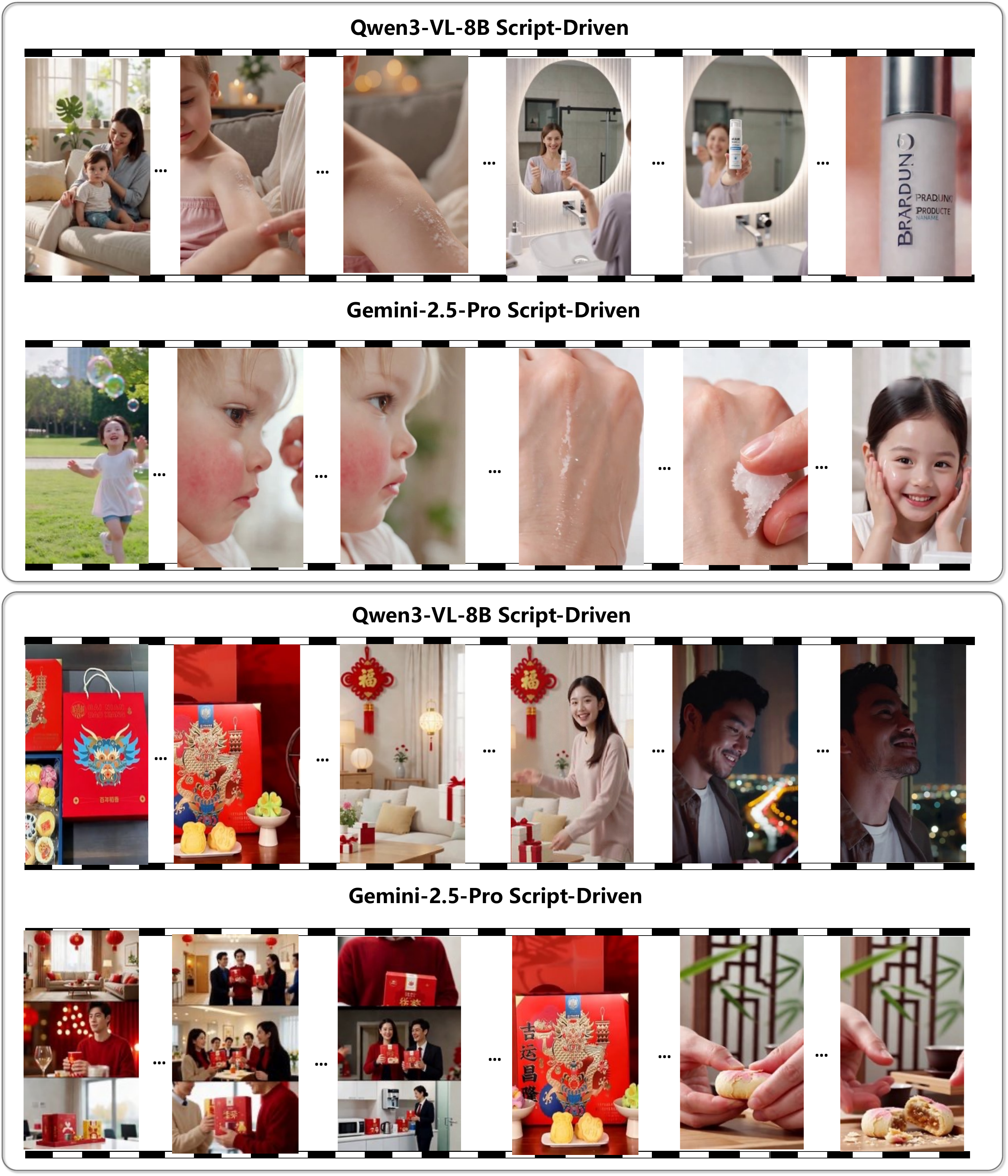} 
    \caption{Qualitative comparison between scripts generated by Qwen3-VL-8B and Gemini-2.5-Pro.}
    \label{fig:video_case2}
\end{figure*}

\begin{figure*}[t]
    \centering
    \includegraphics[width=1.0\linewidth]{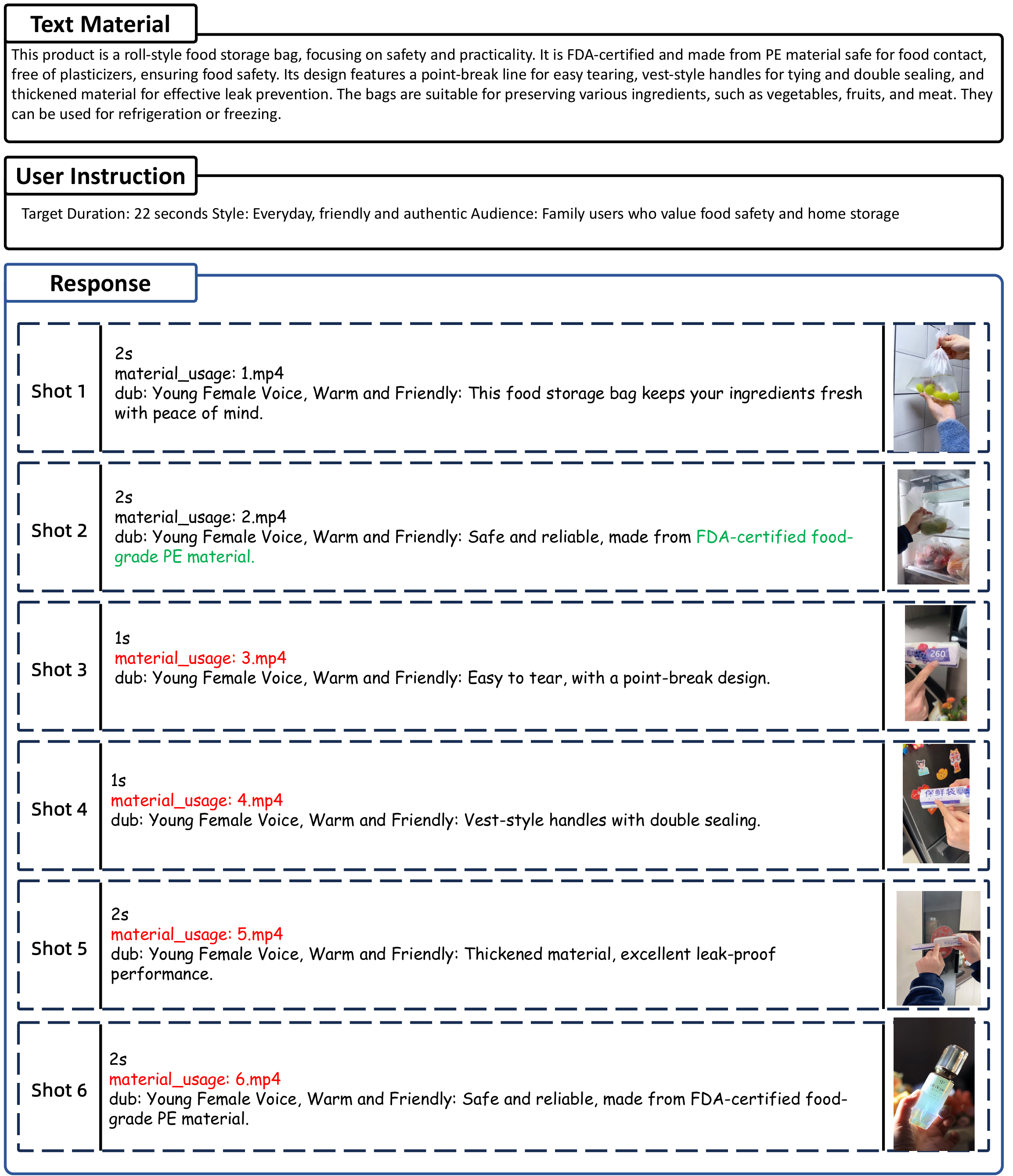}
    \caption{Script generated by Qwen2.5-VL-7B.}
    \label{fig:7b}
\end{figure*}

\begin{figure*}[t]
    \centering
    \includegraphics[width=1.0\linewidth]{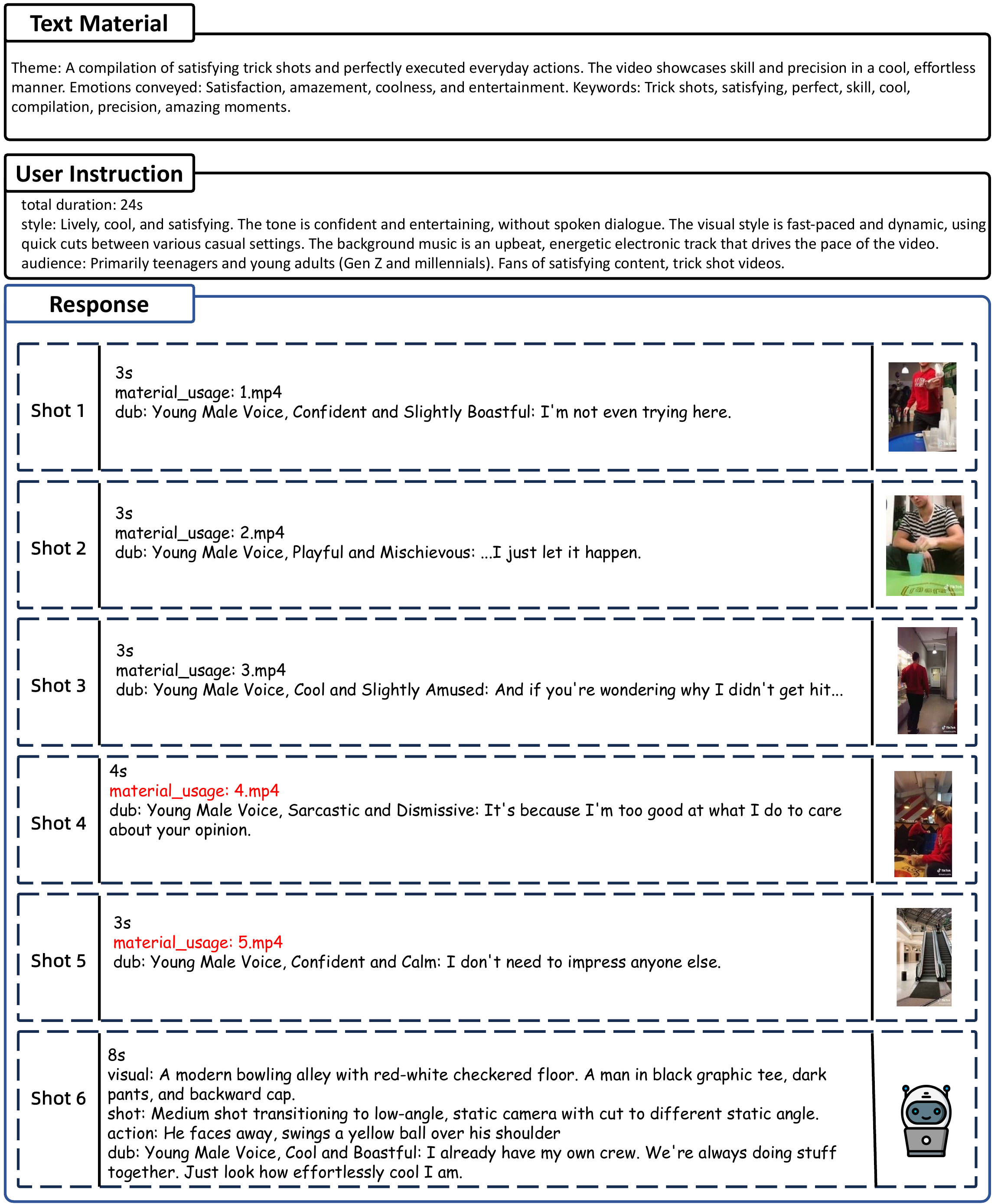}
    \caption{A Skit Script generated by \textit{MCSC-8B}.}
    \label{fig:mcsc8b_gen}
\end{figure*}

\begin{figure*}[t]
    \centering
    \includegraphics[width=1.0\linewidth]{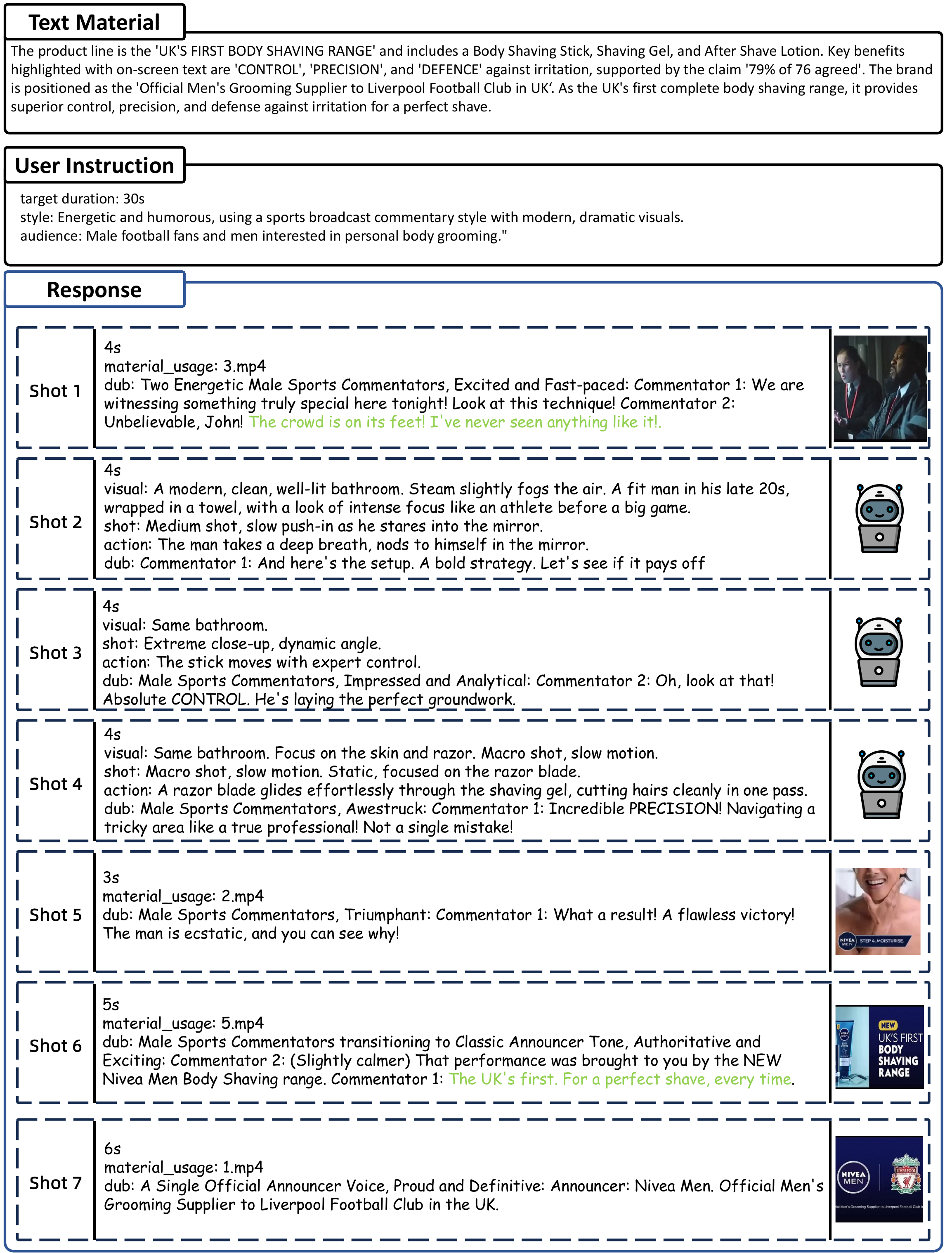}
    \caption{Script generated by Gemini-2.5-Pro.}
    \label{fig:case_gemini_en_ad}
\end{figure*}

\begin{figure*}[t]
    \centering
    \includegraphics[width=1.0\linewidth]{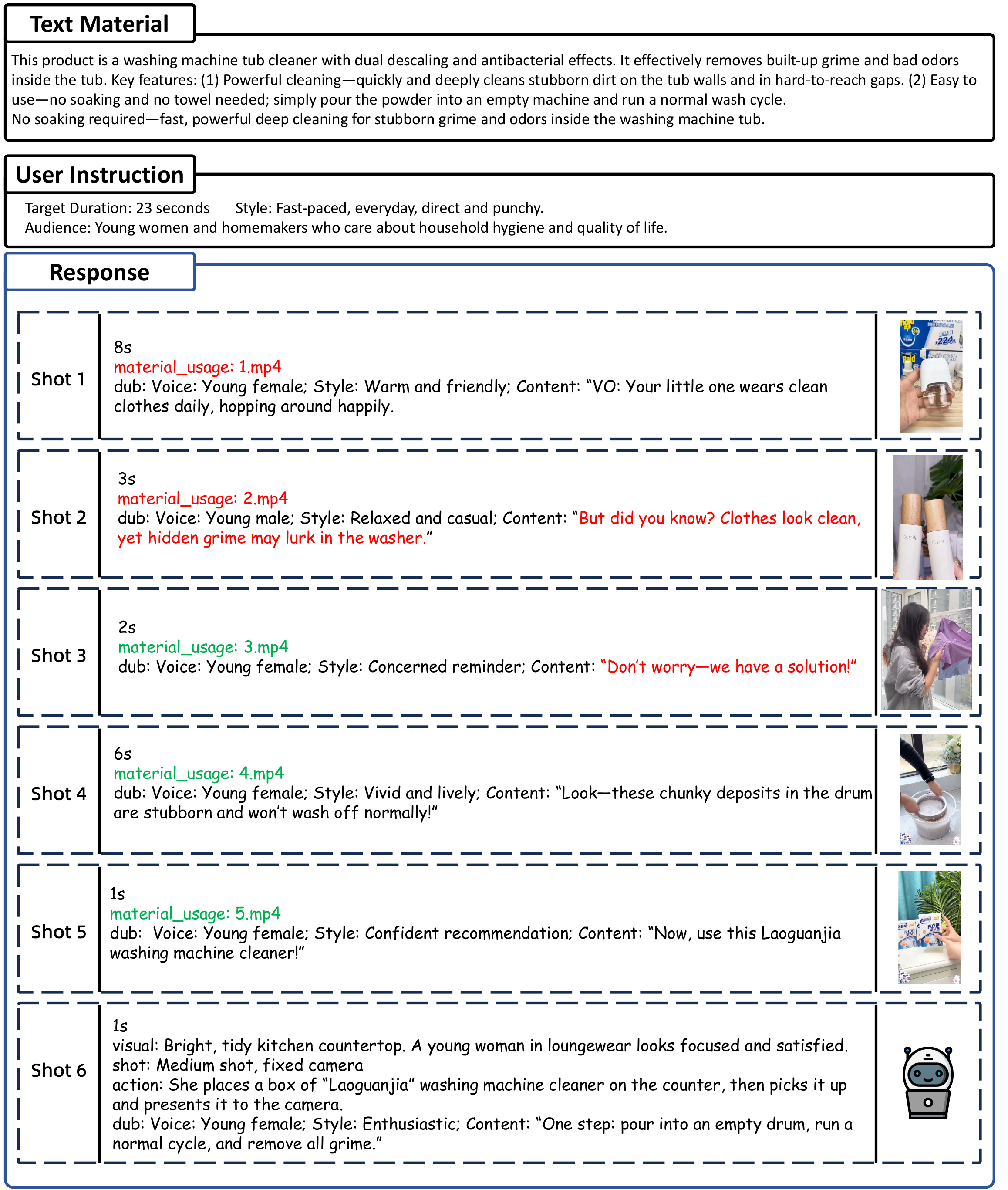}
    \caption{Script generated by Qwen3-VL-8B.}
    \label{fig:8b}
\end{figure*}

\begin{figure*}[t]
    \centering
    \includegraphics[width=1.0\linewidth]{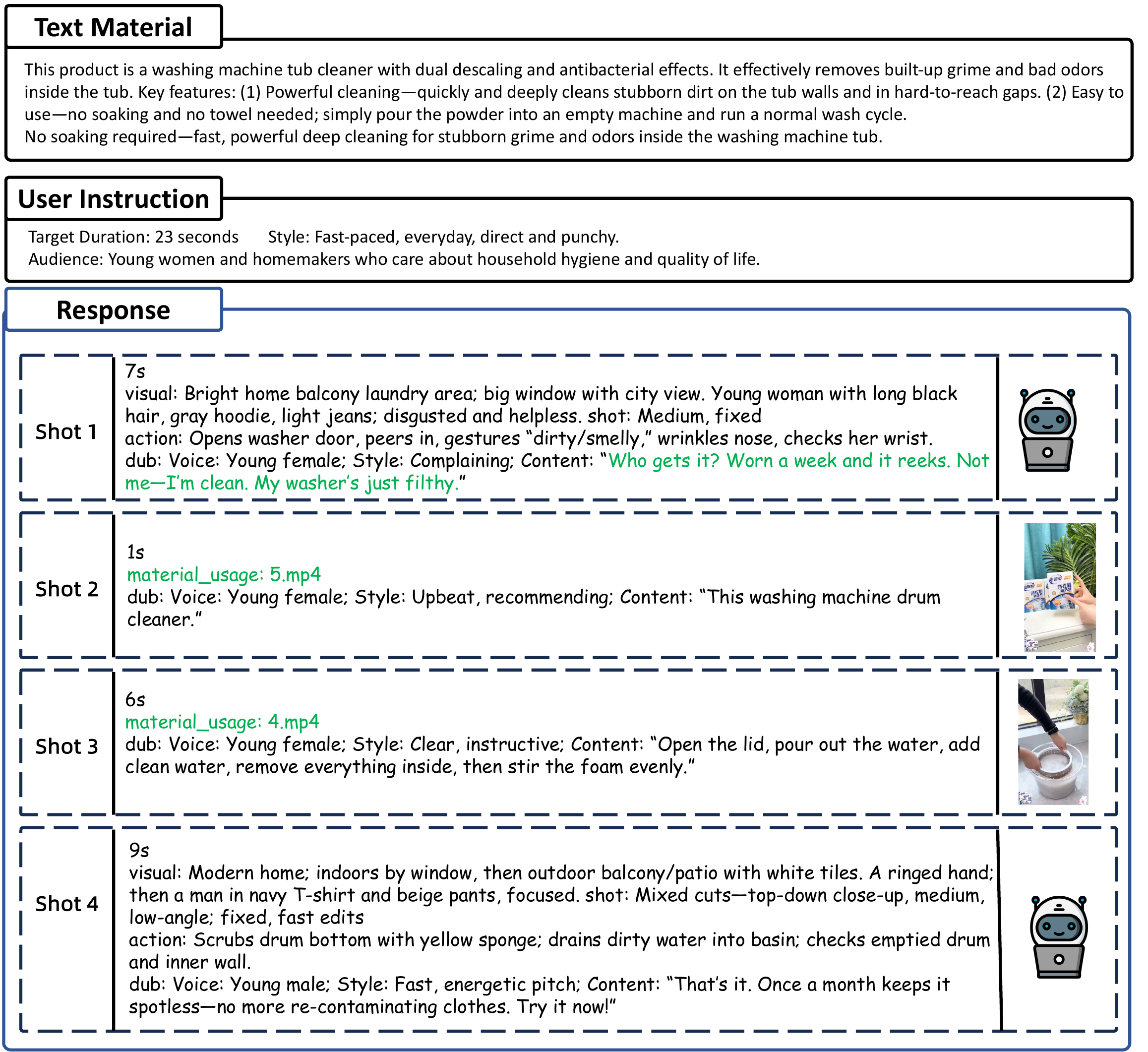}
    \caption{Script generated by \textit{MCSC-8B}.}
    \label{fig:8b-sft}
\end{figure*}

\begin{figure*}[t]
    \centering
    \includegraphics[width=1.0\linewidth]{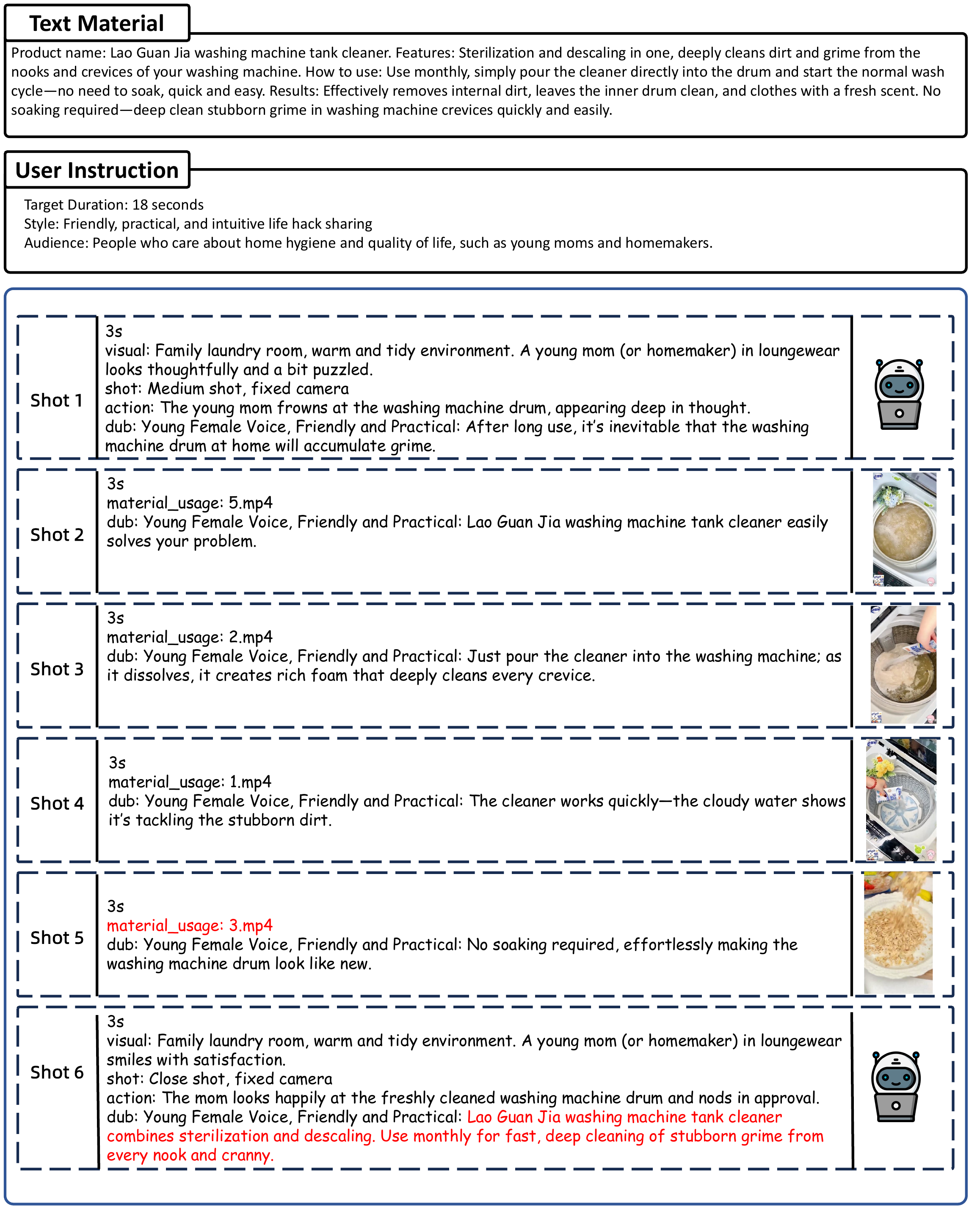}
    \caption{Script generated by Qwen2.5-VL-72B with our agent method.}
    \label{fig:72b-agent}
\end{figure*}

\begin{figure*}[t]
    \centering
    \includegraphics[width=1.0\linewidth]{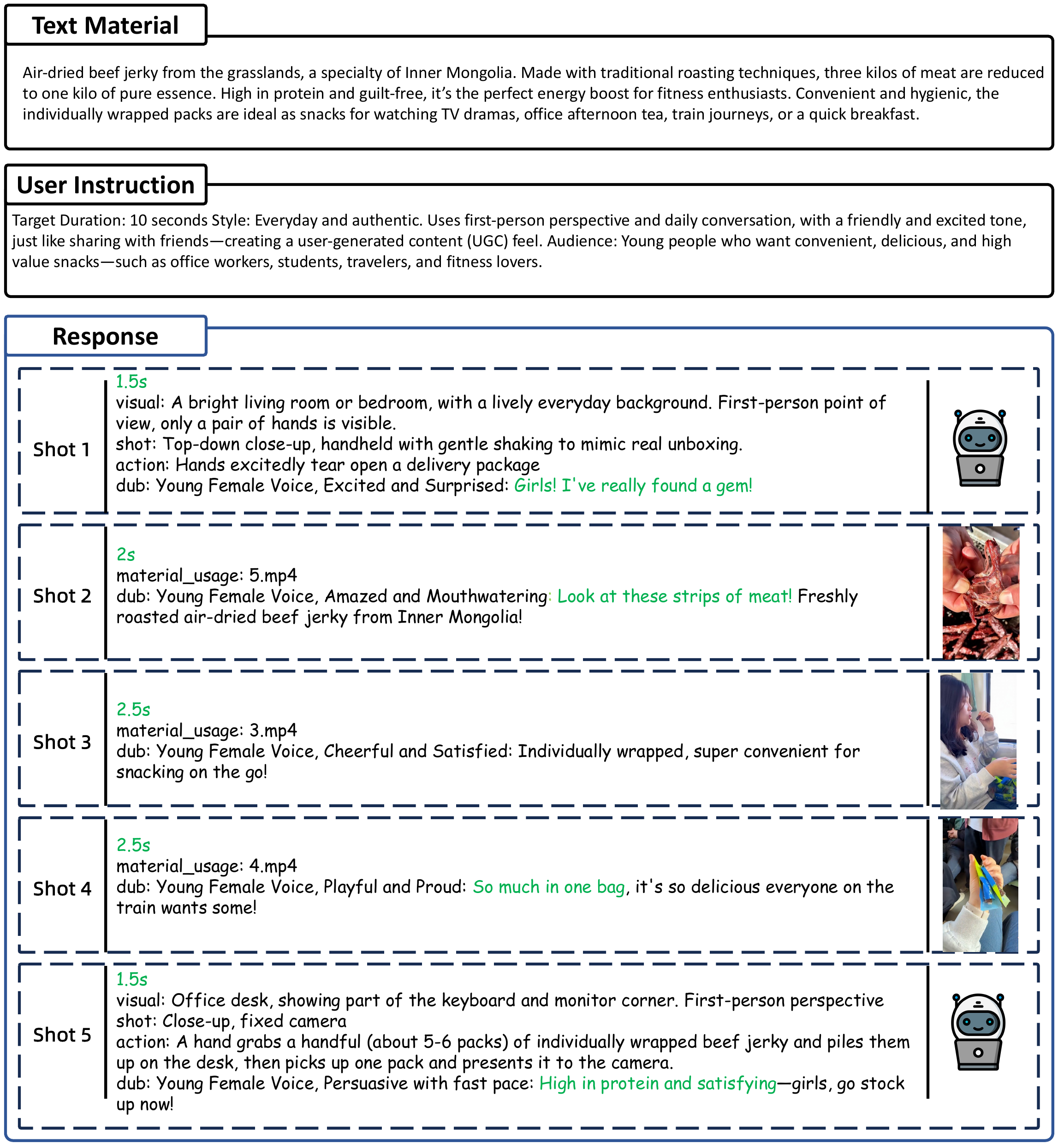}
    \caption{Script generated by Gemini-2.5-Pro.}
    \label{fig:gemini}
\end{figure*}

\begin{figure*}[t]
    \centering
    \includegraphics[width=1.0\linewidth]{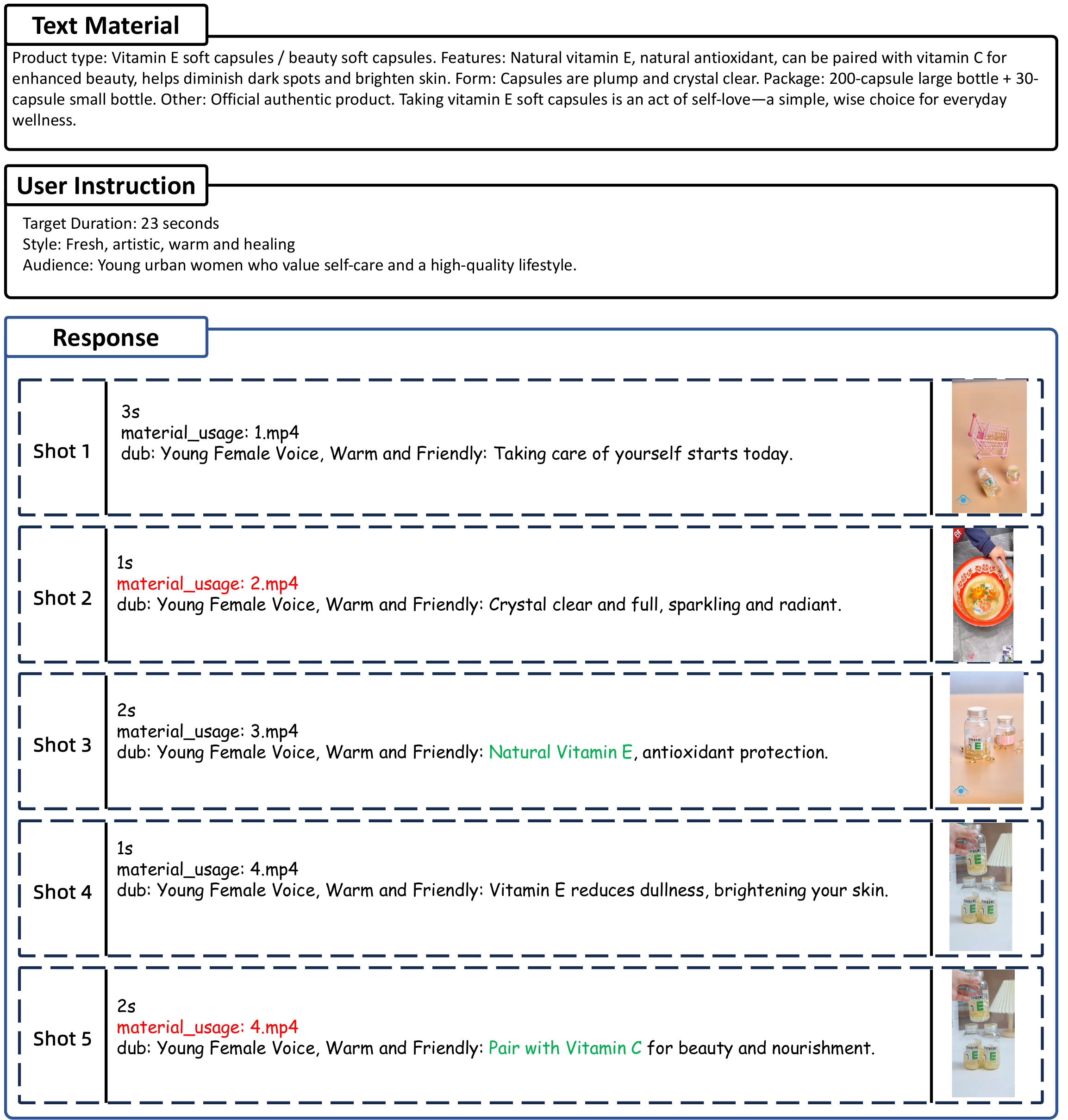}
    \caption{Script generated by InternVL3-8B.}
    \label{fig:intern}
\end{figure*}

\begin{figure*}[t]
    \centering
    \includegraphics[width=1.0\linewidth]{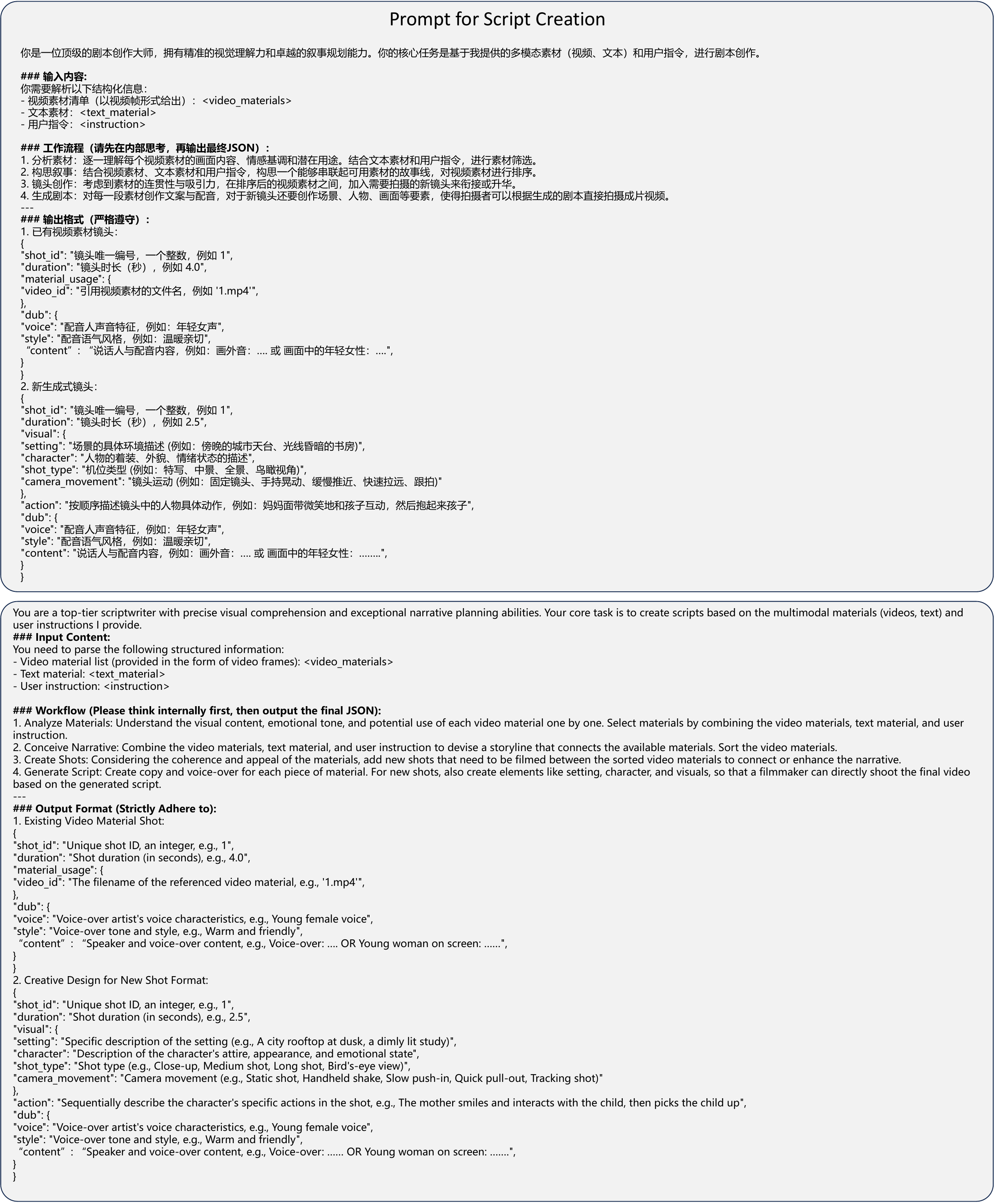}
    \caption{Prompt for script creation in the MCSC task.}
    \label{fig:prompt_compose}
\end{figure*}

\begin{figure*}[t]
    \centering
    \includegraphics[width=1.0\linewidth]{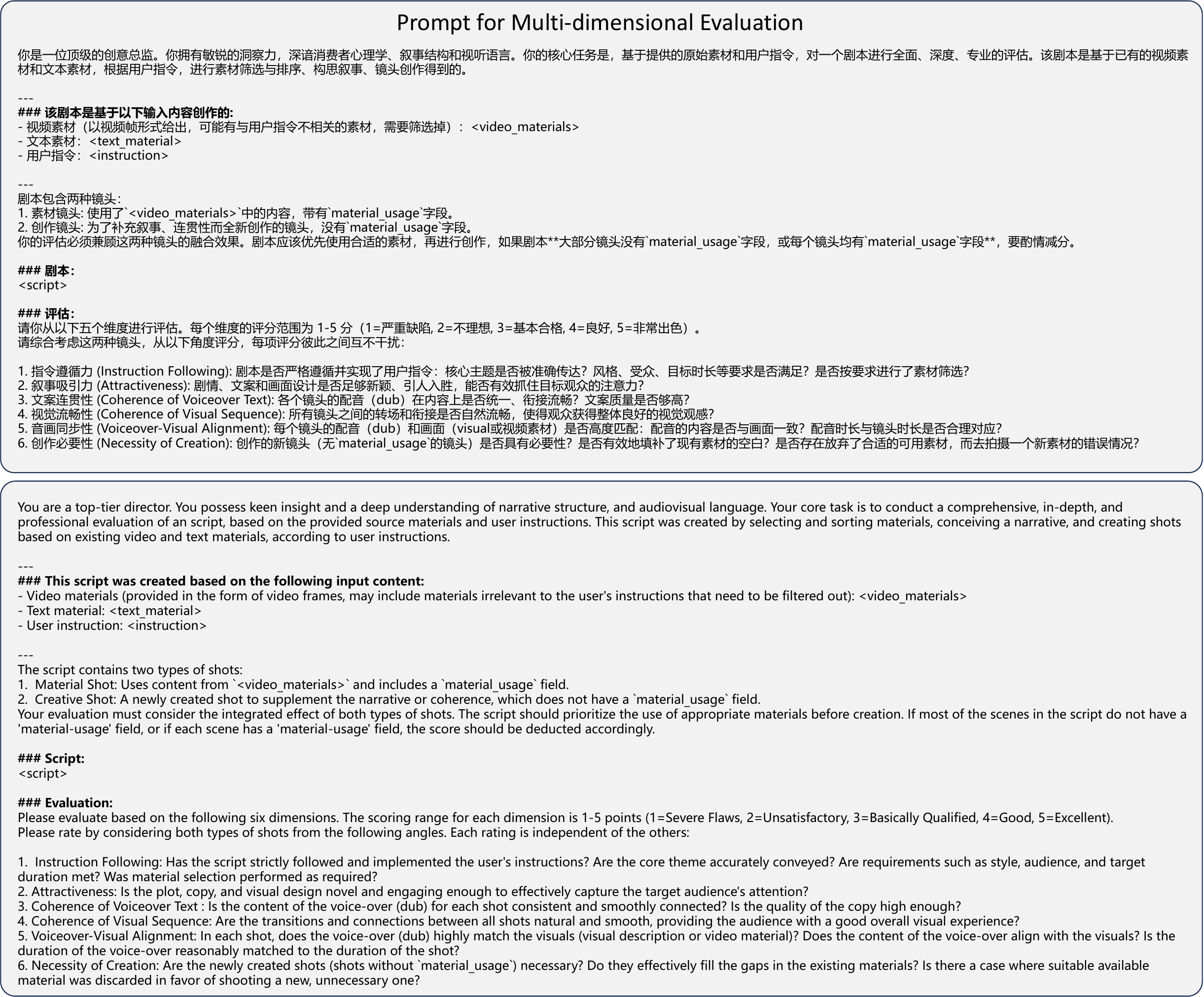}
    \caption{Prompt for script evaluation in the Multi-dimensional metrics(Section~\ref{sec:multi_aspect}). }
    \label{fig:prompt_evaluation}
\end{figure*}

\begin{figure*}[t]
    \centering
    \includegraphics[width=1.0\linewidth]{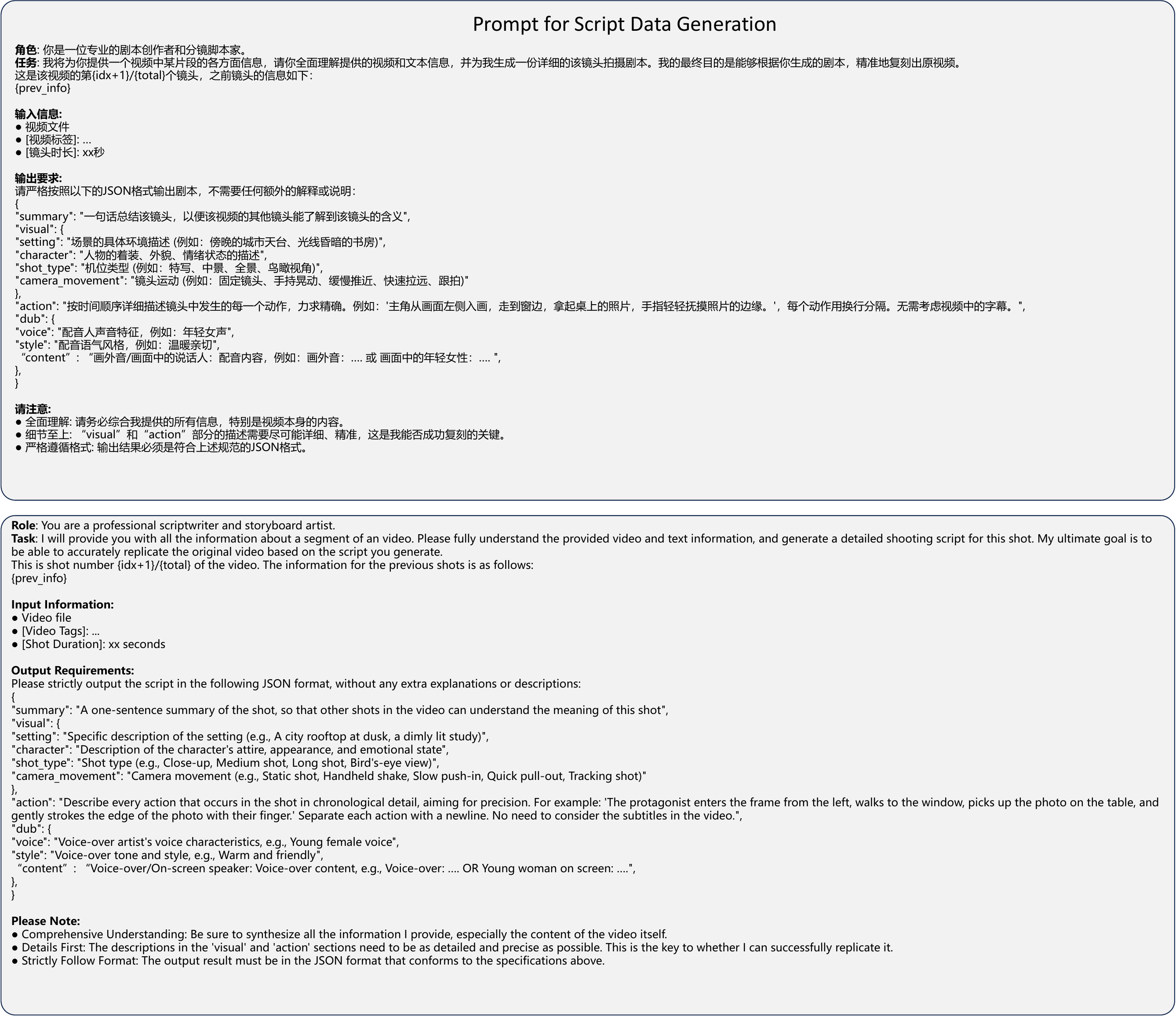}
    \caption{Prompt for script generation in the first phase of Section~\ref{sec:data_pipeline}.}
    \label{fig:prompt_script}
\end{figure*}